\documentclass[aps,manuscrirpt]{revtex4-2} 
\usepackage{bm}
\usepackage{amsmath}
\usepackage{color}
\usepackage{graphicx}
\usepackage{amssymb}
\DeclareMathAlphabet{\mathbbmsl}{U}{bbm}{m}{sl}
\makeatletter
\newsavebox{\@brx}
\newcommand{\llangle}[1][]{\savebox{\@brx}{\(\m@th{#1\langle}\)}%
	\mathopen{\copy\@brx\kern-0.5\wd\@brx\usebox{\@brx}}}
\newcommand{\rrangle}[1][]{\savebox{\@brx}{\(\m@th{#1\rangle}\)}%
	\mathclose{\copy\@brx\kern-0.5\wd\@brx\usebox{\@brx}}}
\makeatother
\begin{document}
\draft

\title{Generalized Peierls substitution for the tight-binding model of twisted multilayer graphene in a magnetic field}

\author{Thi-Nga Do$^{1}\footnote{T.N.Do and P.H.Shih contributed equally}$, Po-Hsin Shih$^{1}$, Hsin Lin$^{2}$, Danhong Huang$^{3}$, Godfrey Gumbs$^{4}\footnote{Corresponding author: {\em E-mail}: ggumbs@hunter.cuny.edu}$, Tay-Rong Chang$^{1}\footnote{Corresponding author: {\em E-mail}: u32trc00@phys.ncku.edu.tw}$}
\affiliation{$^{1}$ Department of Physics, National Cheng Kung University, Tainan 701, Taiwan  \\
$^{2}$ Institute of Physics, Academia Sinica, Taipei 11529, Taiwan \\
$^{3}$US Air Force Research Laboratory, Space Vehicles Directorate (AFRL$/$RVSU),\\
Kirtland Air Force Base, New Mexico 87117, USA \\
$^{4}$ Department of Physics and Astronomy, Hunter College of the City University of New York,
695 Park Avenue, New York, New York 10065, USA}

\date{\today}

\begin{abstract}
	
We propose a generalized Peierls substitution method in conjunction with the tight-binding model to explore the magnetic quantization and quantum Hall effect in twisted multilayer graphene under a magnetic field. The Bloch-basis tight-binding Hamiltonian is employed for large twist angle while an effective tight-binding model is constructed for the magic angle. We investigate extensively the band structures, Landau levels (LLs), and quantum Hall conductivity (QHC) of twisted bilayer graphene and twisted double bilayer graphene, as well as their dependence on the twist angle. Comparison between these crucial properties of monolayer graphene, Bernal bilayer graphene, and the twisted systems is carefully made to highlight the roles played by twisting. The unique selection rules of inter-LL transition, which is crucial for achieving a deep understanding of the step structures of QHC, are identified through the properties of LL wave functions. Our theoretical model opens up an opportunity for comprehension of the interplay between an applied magnetic field and the twisting effect associated with multilayer graphene. The proposed method is expected to be extendable for the calculation of magnetic quantization problems of other complex systems.

\end{abstract}
\maketitle

\section{Introduction}
\label{sec1}

Recently, twisted multilayer graphene (TMLG) has attracted widespread attention since it plays a key role in deciphering the nature of correlation phenomena in low-dimensional materials. Up to now, the most studied TMLG, namely twisted bilayer graphene (TBLG) and twisted double bilayer graphene (TDBLG), have been synthesized successfully by nanotechnology fabrication methods \cite{synthesis}. Notably, TMLG exhibit various intriguing physical properties, such as superconductivity at the magic angle\,\cite{magic1, magic2, magic3}, reentrant correlated insulating states in strong flux of the magic angle TBLG \cite{Bernevig2021}, the nontrivial topology of flat bands\,\cite{topo1, topo2}, the anomalous Hall effect\,\cite{ahe1, ahe2}, the unique magnetic quantization\,\cite{filling1, filling2, double5, LLexp1, LLexp2, double5, LLconti1, LLconti2} and the quantum Hall effect (QHE)\,\cite{PRB2012Moon, PRB2012Fal, PRL2011Lee, PRL2012San, PRL2019Burg, PRB2020Crosse, PRL2021Wu}. These make TMLG a prominent candidate for potential applications in new-generation devices with new advanced functionalities\,\cite{app}.
\medskip

Many fundamental properties of TBLG and TDBLG have been investigated intensively by both theoretical and experimental approaches. Their band structures have been presented from a theoretical perspective by using density-functional theory (DFT)\,\cite{DFT1, DFT2} and modeling\,\cite{Mac, Koshino, Fang, effective, tbm1, tbm2, tbm3, double1, double2}. Additionally, band structures have also been obtained experimentally by utilizing scanning-tunneling microscopy (STM) and spectroscopy (STS) techniques\,\cite{double5, bandexp1, bandexp2, bandexp3, bandexp4, bandexp5, double2, double3, double4}. Besides electronic structures, the magnetic quantization which provides important dynamical information of materials has been identified experimentally\,\cite{LLexp1, LLexp2} and predicted theoretically\,\cite{double5, LLconti1, LLconti2} for TBLG and TDBLG. It has been shown for TBLG that its quantum Hall conductivity (QHC) exhibits a step structure, in which the plateau series (in the unit $e^2/h$) are $\pm4$, $\pm8$, $\pm12\cdots$\,\cite{PRB2012Moon, PRB2012Fal, PRL2011Lee, PRL2012San}. Meanwhile, the studies for QHE in TDBLG with small twist angles have also been carried out. However, their filling factors are responsive to the stacking configuration and magnetic field\,\cite{PRL2019Burg, PRB2020Crosse, PRL2021Wu}.
Though remarkable efforts have been devoted to studies of the transport properties of TMLG, a deep understanding of the QHC step formation, as well as its connection to the quantized Landau levels (LLs), still remain a mystery. The limitation of theoretical study on TMLG results from the complexity of atomic-scale structures.
A suitable tight-binding model (TBM) for the TMLG is thus highly desirable.
\medskip

In this paper, we comprehensively investigate the magneto-electronic and QHE of TMLG and their dependence on the twist angle by using the Bloch-basis TBM in conjunction with the generalized Peierls substitution. This method allows us not only to obtain accurate band structures in a wide energy range but also to solve the huge Hamiltonian matrix under a magnetic field. Specifically, the calculated LL wave functions enable a full analysis of transition-selection rules which are very useful for an extensive understanding of the QHE. We will present the band structures, field-dependent LL spectra, LL wave functions, and Fermi energy-dependent QHC. We will also show these characteristics for monolayer graphene (MLG) and AB bilayer graphene (BLG) which are critical for analyzing and understanding our numerical results. Comparison between the TMLG and these two systems will be made to clarify the principle role played by the twist angle in electronic and transport properties.
It is critical to mention that our calculated results of LL spectra and filling factors for both large twist angle and small magic angle are in good agreement with the experimental measurements.
\medskip

The rest of the paper is organized as follows. In Sec.\,\ref{sec2}, we first establish our theoretical model for executing numerical computations, including tight-binding formalism, application of a magnetic field, effective TBM and Kubo formula. In Sec.\,\ref{sec3}, we present a detailed discussion on obtained numerical results for considered structures. Finally, conclusions drawn from this paper are summarized in Sec.\,\ref{sec4}.

\section{Theoretical Method}
\label{sec2}

In this work, the $\pi$-electronic structure is calculated using the $p_z $-orbital TBM. The Kubo formula is combined with the TBM to evaluate the Fermi energy-dependent QHC. Some details are presented below.

\subsection{Tight-binding model}

\begin{figure}[htbp]
\begin{center}
\includegraphics[width=0.75\linewidth]{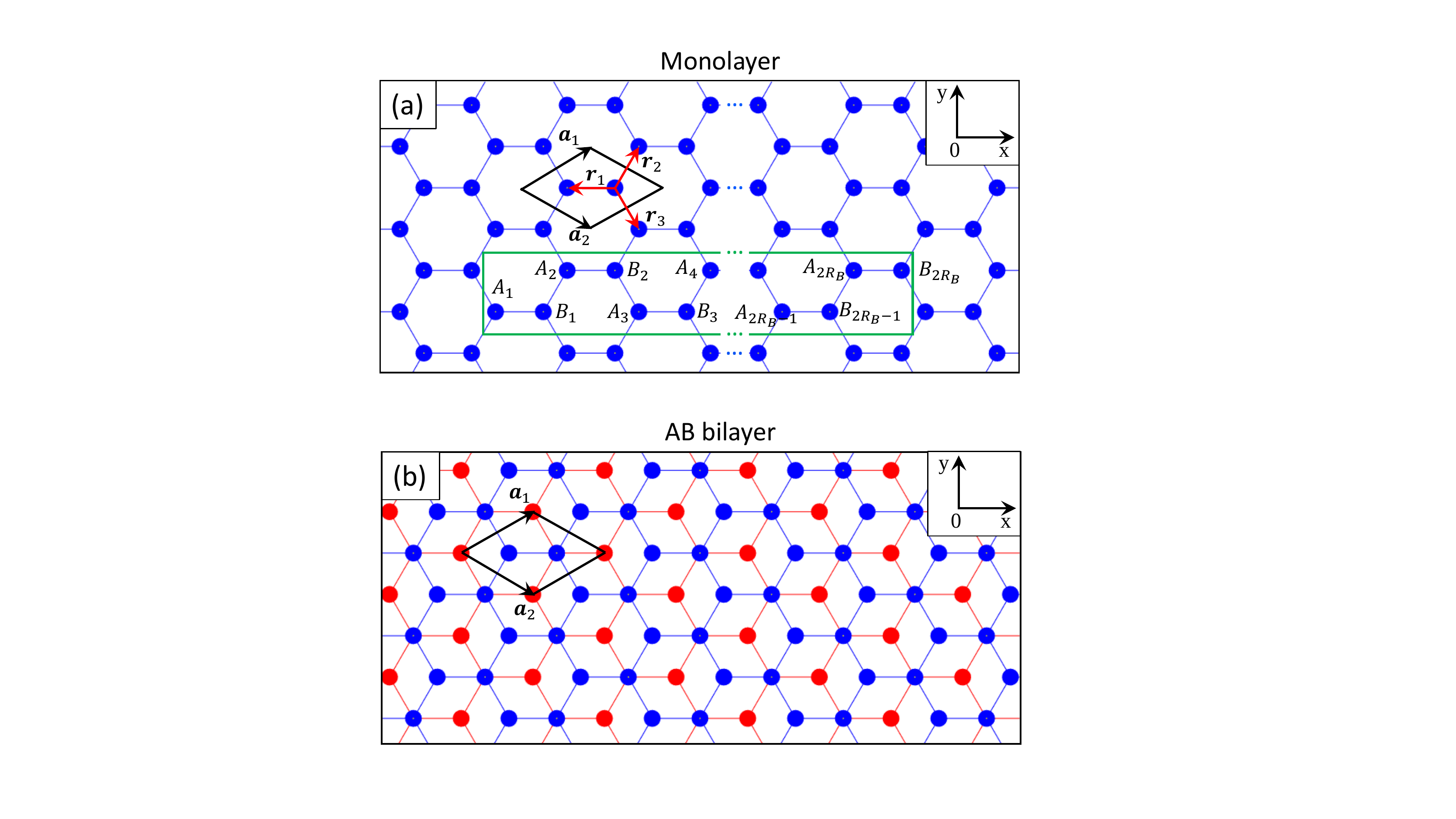}
\end{center}
\caption{(color online) Lattice structure of $(a)$ monolayer graphene and $(b)$ AB bilayer graphene. The unit cell is outlined by black lines, $\mbox{\boldmath$a$}_1$ and $\mbox{\boldmath$a$}_2$ are the lattice vectors. [$\mbox{\boldmath$r$}_1$, $\mbox{\boldmath$r$}_2$, $\mbox{\boldmath$r$}_3$] are three nearest-neighbor vectors. The magnetic-field enlarged unit cell is outlined by green lines in $(a)$. The red and blue balls indicate atoms on the first and second layers, respectively.}
\label{Fig1}
\end{figure}

For monolayer graphene, there are two equivalent sublattices designated as $A$ and $B$. The primitive unit cell consists of two atoms, as illustrated in Fig.\,\ref{Fig1}(a). Here, $\mbox{\boldmath$a$}_1$ and $\mbox{\boldmath$a$}_2$ are the lattice vectors, $a = |\mbox{\boldmath$a$}_1| = |\mbox{\boldmath$a$}_2| \approx 2.46\,$\AA\ is the lattice constant. For Bernal bilayer graphene, the two graphene sheets are separated by the interlayer distance of $d_0 \approx 3.35\,$\AA. The notations $A^{\ell}$ and $B^{\ell}$ indicate two sublattices on the $\ell$-th $(\ell=1,2)$ layer (Fig.\,\ref{Fig1}(b)). There are four carbon atoms in a primitive unit cell. The $(x,y)$ coordinates of two layers are different by a shift of a C-C bond length ($a_0= 1.42\,$\AA) along the armchair direction. The tight-binding Hamiltonian for a layered graphene system can be expressed as

\begin{equation}
H  = \sum\limits_{m,j;\,\ell,\ell^{\prime}} t_{mj}^{\ell\ell^{\prime}}(c_{m}^{\ell})^\dagger c_{j }^{\ell^{\prime}} + H.c.\ .
\label{eqn:1}
\end{equation}
In Eq.\,\eqref{eqn:1}, $m$ and $j$ denote the lattice sites, $c_{m}^{\ell}$ [$(c_{m}^{\ell})^\dagger$] is the annihilation (creation) operator which can destroy (generate) an electronic state at the $m$-th site on the $\ell$-th layer, $t_{mj}^{\ell\ell^{\prime}}$ are the hopping interaction terms between the atoms at the $m$-th site on the $\ell$-th layer and the $j$-th site on the $\ell^{\prime}$-th layer, and H.c. stands for Hermitian conjugation.
\medskip

The hopping interaction between two C atoms at lattice points $\mbox{\boldmath$R$}_m$ and $\mbox{\boldmath$R$}_j$ is calculated as

\begin{equation}
-t_{mj} = V_{pp\pi} \left [ 1- \left( \frac{\mbox{\boldmath$h$}\cdot\mbox{\boldmath$e$}_z}{h}   \right )^2  \right ] +  V_{pp\sigma} \left( \frac{\mbox{\boldmath$h$}\cdot\mbox{\boldmath$e$}_z}{h}   \right )^2,
\label{eqn:2}
\end{equation}
where

\begin{eqnarray}
\nonumber
V_{pp\pi}&=&V_{pp\pi}^0\,\exp\left( -\frac{h-a_0}{\delta}   \right )\ ,\\
\nonumber
V_{pp\sigma}&=&V_{pp\sigma}^0 \,\exp\left( -\frac{h-d_0}{\delta}   \right )\ .
\end{eqnarray}
Here, $\mbox{\boldmath$e$}_z$ is the unit vector in the $z$ direction, $V_{pp\pi}^0$ is the nearest-neighbor transfer integral in monolayer graphene ($t_{mj}^{\ell\ell}$ in Table I), $V_{pp\sigma}^0$ is the interlayer transfer integral between vertical atoms. Also, $\delta = 0.184\,a$ is the decay length of transfer integral. The TBM in this work includes hopping interactions within the area of $h \leq 4a_0$. The parameters for hopping interactions of monolayer graphene and Bernal bilayer graphene are presented in Table I.
\medskip

\begin{table}[ht]
\caption {Hopping parameters of monolayer graphene and AB bilayer graphene}
\begin{tabular}{|c|c|c|}
\cline{1-3}
Hopping parameters [eV]& Monolayer & AB bilayer   \\ \cline{1-3}
$t_{mj}^{\ell\ell}$ &$-2.7$ & $-2.7$  \\ \cline{1-3}
$t_{mm}^{\ell\ell} (B)$ & $0$ & $0.0366$    \\ \cline{1-3}
$t_{m,m}^{12}$ & $0$ & $0.48$   \\ \cline{1-3}
$t_{m,m\pm1}^{12}$ & $0$ & $0.211$  \\ \cline{1-3}
\end{tabular}
\label{tab1}
\end{table}

Solving the Hamiltonian is of utmost importance for gaining an understanding of the essential physical properties of the materials. Due to the limitation of the numerical technique, constructing a sufficiently small tight-binding matrix Hamiltonian for the large systems is highly desirable. So far, several methods have been widely employed for reducing the size of the Hamiltonian matrix, such as the scalable TBM \cite{Liu}, the continuum model \cite{Mac, Koshino}, and the effective TBM (minimal model) \cite{Fang, effective}. The scalable TBM for graphene has been built up with the scaled hopping parameter and lattice spacing. This model describes the so-called ``theoretical artificial graphene" which has been proved to capture the same results for the electronic and transport properties as graphene. This method is restricted to the long wavelength limit for which the Fermi wavelength should be much longer than the lattice spacing. In the presence of a magnetic field, the validity of the Peierls substitution imposes a further restriction for the scaling, in which the magnetic length must be much larger than the lattice spacing.  On the other hand, the continuum model and minimal model are limited to low-energy physics. In addition, these models cannot capture the boundary conditions of systems. Therefore, some essential physical features might be excluded.  Here, we show that our proposed Bloch-based TBM can deal with larger systems and in a wider range of energy. Furthermore, the combination of the Bloch function base and the effective TBM enables the investigation of fundamental properties of TMLG at the magic angle under a magnetic field.

\medskip

It is well known that the Wannier and Bloch function bases are commonly used to construct the TBMs of graphene and other condensed matter systems. The crucial difference between these two results from the geometric phases of interacting atoms. For the Wannier basis, the distance between two atoms is estimated based on their Cartesian coordinates. For the Bloch basis, on the other hand, a unit cell is viewed as a point so that the distance between any two atoms is just the distance between the two unit cells which they belong to. For monolayer graphene, three nearest-neighbor geometric phases (see Fig.\,\ref{Fig1}(a)) can be expressed either by the Wannier basis [$\mbox{\boldmath$r$}_1(B) = (-b, 0)$; $\mbox{\boldmath$r$}_2(B) = (b/2, -\sqrt{3}b/2)$; $\mbox{\boldmath$r$}_3(B) = (-b/2, \sqrt{3}b/2)$]  or by the Bloch basis [$\mbox{\boldmath$r$}_1(B) = (0, 0)$; $\mbox{\boldmath$r$}_2(B) = (3b/2, \sqrt{3}b/2)$; $\mbox{\boldmath$r$}_3(B) = (3b/2, -\sqrt{3}b/2)$]. These vectors are associated with geometric phases of different lattice sites. Our numerical results of band structures for monolayer graphene and AB bilayer graphene by using TBMs based on Wannier and Bloch bases agree with each other.
In general, the Bloch function basis is advantageous over the Wannier function one for large systems (with many atoms within a unit cell) since the former simplifies the Hamiltonian matrix by reducing the number of geometric phases.
\medskip

\begin{figure}[htbp]
\begin{center}
\includegraphics[width=0.6\linewidth]{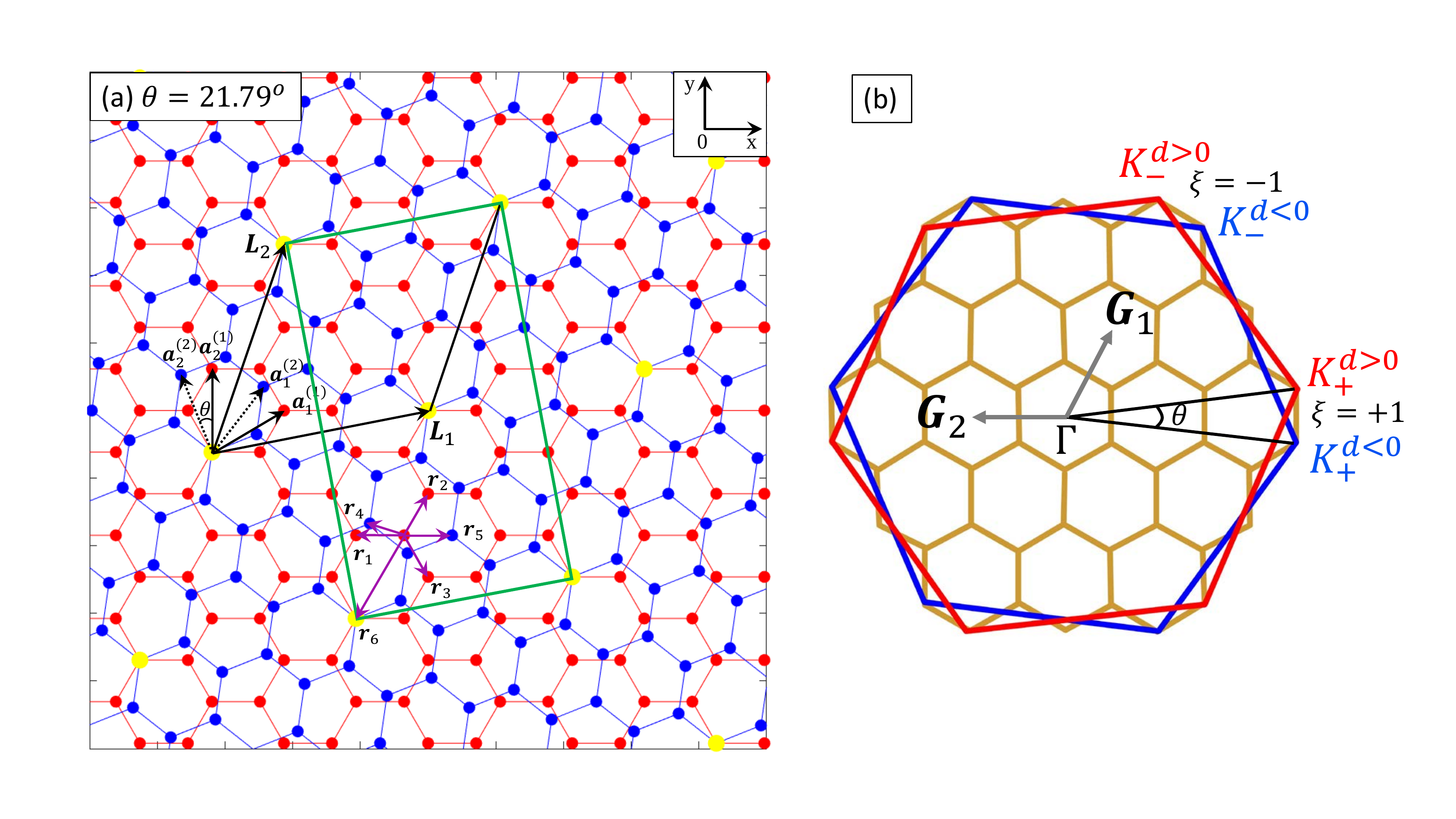}
\end{center}
\caption{(color online) (a) Lattice structure of a twisted bilayer graphene with a twisted angle $\theta = 21.79^{\rm o}$. The blue and red balls denote atoms on the bottom and top layers, respectively. The overlapped atoms are indicated by yellow balls. The unit cell is outlined by the black lines. [$\mbox{\boldmath$r$}_1$, $\mbox{\boldmath$r$}_2$, $\mbox{\boldmath$r$}_3$, $\mbox{\boldmath$r$}_4$, $\mbox{\boldmath$r$}_5$, $\mbox{\boldmath$r$}_6$] are selected nearest-neighbor vectors. [$a_1^{(1)}$, $a_2^{(1)}$] and [$a_1^{(2)}$, $a_2^{(2)}$] denote the lattice vectors of the first and second layers, respectively; $L_1$ and $L_2$ are the lattice vectors of TBG. (b) shows the selected first Brilloun zones for the bottom (blue) and top (red) layers, where [$K_+^{d<0}$, $K_-^{d<0}$] and [$K_+^{d>0}$, $K_-^{d>0}$] represent two valleys of the bottom and top layers, respectively.}
\label{Fig2}
\end{figure}

For TBLG, two graphene layers are relatively rotated by an angle called  the ``twist angle''. Figure\,\ref{Fig2}(a) presents the lattice structure of TBLG with a twist angle $\theta = 21.79^{\rm o}$. The lattice vectors of the TBLG [$\mbox{\boldmath$L$}_1$ and $\mbox{\boldmath$L$}_2$] can be expressed in terms of the lattice vectors of the first [$\mbox{\boldmath$a$}_1^{(1)}$, $\mbox{\boldmath$a$}_2^{(1)}$] and second [$\mbox{\boldmath$a$}_1^{(2)}$, $\mbox{\boldmath$a$}_2^{(2)}$] layers. Specifically, $\mbox{\boldmath$L$}_1 = m\mbox{\boldmath$a$}_1^{(1)} + n \mbox{\boldmath$a$}_2^{(1)} = n\mbox{\boldmath$a$}_1^{(2)} + m \mbox{\boldmath$a$}_2^{(2)}$ and $\mbox{\boldmath$L$}_2 = R(\pi/3)\,\mbox{\boldmath$L$}_1$, where $m$ and $n$ are certain integers satisfying $\theta(m,n) = arg[(me^{-i\pi/6} + ne^{i\pi/6})/(ne^{-i\pi/6} + me^{i\pi/6})]$, and $R(\theta)$ defines the rotation by an angle $\theta$\,\cite{twistmodel, Koshino2013}. Figure\,\ref{Fig2}(b) displays the Brillouin zones for the first (red hexagon) and second (blue hexagon) layers. The Brillouin zone of TBLG is also a hexagon but with a smaller size compared to Brillouin zones of two individual layers.
For a TBLG with a large twist angle, the two layers are considered decoupled so that it can be described by the TBM of BLG in the absence of interlayer interactions.
\medskip

We note that TDBLG will acquire various stacking configurations since the BLG components can be AA or AB types. Here, we consider  twisted double AB-AB bilayer graphene due to its special lattice symmetry. This system consists of a pair of Bernal BLG with a relative rotation between them. Each AB BLG contains the atomic interactions as quantified in Table\,\ref{tab1}. Furthermore, the middle two layers resemble the TBLG. Interestingly, there exists an overlapping behavior of certain atoms on different graphene layers, e.g., the mutual coordinates of ($A^1$, $A^2$, $A^3$, $A^4$), ($A^1$, $A^2$), ($A^3$, $A^4$), ($B^1$, $B^4$) and ($B^2$, $B^3$). In our calculations, only the atomic interactions between the nearest-neighbor graphene layers are taken into account. Therefore, the tight-binding Hamiltonian of the AB-AB TDBLG can be constructed based on those of the AB BLG and TBLG.

\subsection{Generalized Peierls substitution}

As a graphene sheet is subjected to a uniform perpendicular magnetic field $\mbox{\boldmath$B$}= (0,0,B)$, the field-induced Peierls phase $G_{R}$ needs to be included in the graphene Hamiltonian. The Peierls substitution\,\cite{Pei1933} is a common method to study the behavior of Bloch electrons in a magnetic field. This approach can be classified into two mechanisms: (1) The substitution of $-i\hbar \nabla -\frac{e}{c} \mbox{\boldmath$A$}$ for $\hbar \mbox{\boldmath$k$}$ in the energy function for a band\,\cite{Hof1976} and (2) the multiplication of the zero-field matrix elements of the tight-binding Hamiltonian by the Peierls phase factors\,\cite{Ezawa2008, Bernevig2020}.
It is noted that, the approach (1) is limited to the simple lattices for which the expression of energy function can be accomplished, e.g, the square lattice\,\cite{Hof1976}.
On the other hand, the Peierls substitution can be made in the matrix elements of the tight-binding Hamiltonian, according to the approach (2), so that the resulting equations near the K and K$^{\prime}$ points can be expanded to lowest orders in the wave vector and the vector potential. In particular, the Peierls substitution made for the momentum operator of the effective massless Dirac Hamiltonian has been frequently used in the literature\,\cite{Ezawa2008, Bernevig2020}.
However, this procedure is restricted to the low-energy range of the systems with simple band structures.
Here, we propose using the mechanism (2) and numerically solve the Hamiltonian without further perturbation. This method can yield the accurate results of a wide energy range for a large group of materials.
\medskip

The Peierls phase can be expressed in term of the magnetic vector potential $\mbox{\boldmath$A$}$ via the relation $G_{R} = (2\pi/\phi_{0})\int\limits_{\bf R}^{\bf r} \mbox{\boldmath$A$}\cdot d\mbox{\boldmath$\ell$}$, in which $\phi_{0}=hc/e$ is the flux quantum. Without loss of generality, we set $c= 1$ in our calculations. Within the Landau gauge, the vector potential is written as $\mbox{\boldmath$A$}=(0,Bx,0)$. Accordingly, the period of the Peierls phase is defined as $2\phi_{0}/ \phi$, in which $\phi = B{\cal S}$ is the magnetic flux with ${\cal S}$ being the area of a unit cell. The applied magnetic field leads to the extension of a unit cell along the $x$ direction (see Fig.\,\ref{Fig1}(a) for monolayer graphene) so that it will include the number of $2N \times 2\phi_{0}/ \phi$ atoms. Here, $2N$ is the number of atoms in the reduced super cell with $N$ being the number of atoms in the zero-field unit cell.
\medskip

The magnetic Hamiltonian matrix elements can be expressed as

\begin{equation}
H_B =\sum\limits_{m,j;\,\ell,\ell^{\prime}}\, t_{mj}^{\ell\ell^{\prime}}\,e^{iG_R}\,\left(c_{m}^{\ell}\right)^\dagger c_{j}^{\ell^{\prime}} + H.c.\ ,
\label{eqn:3}
\end{equation}
where $\ell,\,\ell'$ and $j,\,m$ are layer and site indexes, respectively, and $H.c.$ represents the Hermitian conjugate term.
In the case of TBLG, the field-induced extension of a unit cell is along the direction of $\mbox{\boldmath$L$}_1$, as seen in Fig.\,\ref{Fig2}(a). The main challenge of theoretical models in studying TMLG under a magnetic field lies in the issue of large number of atoms included within a field-extended unit cell. In order to resolve this issue, we propose to employ the Bloch function basis in  investigating the characteristics of LLs and QHC. Using Bloch basis, we consider only the Peierls phases between the atoms in different unit cells but ignore those within the same unit cell. In this way, it enables the computation of LLs for a huge magnetic Hamiltonian matrix, which becomes impractical by using the Wannier function basis. It is noted that the Peierls phases are added to the hopping terms of Hamiltonian in Eq.\,\eqref{eqn:3}. Our calculated results for monolayer graphene and AB bilayer graphene (see Figs.\,\ref{Fig3} and \ref{Fig4}) are in good agreements with previous experimental and theoretical studies. In connection with the Hamiltonian in Eq.\,\eqref{eqn:3}, the LL wave function $|\psi\rangle$ takes the form

\begin{equation}
|\psi\rangle = \sum\limits_{j=1}^{2\phi_{0}/ \phi}\left({\cal A}_j | A_j \rangle    + {\cal B}_j | B_j \rangle\right),
\label{eqn:4}
\end{equation}
where ${\cal A}_j$ and ${\cal B}_j$ are the subenvelope functions and represent the amplitudes of tight-binding wave functions for $j$-th $A$ and $B$ atoms in a unit cell, respectively,
while $| A_j \rangle$ and $| B_j \rangle$ are their corresponding eigenstates.

\subsection{Effective tight-binding model}

Physically, our proposed Bloch basis used in the TBM for calculating the LLs of TBLG can be extended to other twisted structures with arbitrary twist angle $\theta$, e.g. TDBLG and twisted graphene on substrates. Technically, however, if the twist angle is sufficiently small, e.g. the magic angle, the primitive unit cell becomes sizable so that the field-extended unit cell is too large for numerical computations. In addition, the smaller the twist angle is, the weaker the magnetic field is needed to quantize electronic states into LLs. As a result, a lower limit is expected for a twist angle in studying magnetic quantization with TBM. Here, we propose using an effective four-band TBM for exploring LL dynamics together with QHE in TBLG and TDBLG at the magic angle.
\medskip

We construct an effective four-band TBM for TMG based on the Wannier TBM for few-layer graphene (FLG) and continuum model, following the ref. \cite{effective}.
The low-energy Hamiltonian for TMLG is built up with 2 $\times$ 2 block-diagonal elements\,\cite{Fang}, where each block represents Bloch waves of individual MLG at various momentum states. These block-diagonal elements are coupled to one another through interlayer interactions. Such couplings break down the translation symmetry of graphene unit cell so that the new supercell is required for TMLG. Here, the reciprocal lattice of this supercell is utilized to transform the MLG block from a real-space basis to a momentum basis.
The chiral decomposition has been demonstrated for few-layer graphene (FLG) and TMLG \cite{effective}, in which a general stacking sequence can be structurally decomposed into several partitions with a chirally stacking order. Consequently, the low-energy states of a FLG can be well described by the sum of subspaces in terms of the so-called ``pseudospin doublets''.
For TMLG, these pseudospin doublets include the renormalized four flat bands and the remaining ones.
Let us consider a general TMLG with N top-layers and M-bottom layers with small twist angles $\pm \theta/2$, respectively. The Wannier tight-binding model can be written as
\begin{equation}
H_{mn} ({\bf R}) = \frac{1}{N_{\bf k}}\sum\limits_{\bf k}e^{-i {\bf k}.{\bf R}} \langle \psi_{m {\bf k}}|E_{\bf k}I| \psi_{n {\bf k}}\rangle,
\end{equation}
where $E_{\bf k}$ is the eigenvalue and $\psi_{m {\bf k}}$ is the Bloch sum function. They can be obtained from the effective continuum model
\begin{equation}
\hat{H}({\bf k}) = \hat{H}_0({\bf k}) + \hat{H}_T.
\end{equation}
Here, $\hat{H}_0({\bf k})$ is the Hamiltonian of the FLG in the top and bottom of a TMLG and $\hat{H}_T$ describes the effective twisted interlayer coupling.

\medskip
The initial guess for the Bloch sum functions is
\begin{equation}
|\psi^{(0)}_{n{\bf k}} \rangle = \sum\limits_{m}| \Psi_{m{\bf k}}\rangle \langle \Psi_{m{\bf k}} | g_n \rangle,
\end{equation}
in which, $|\Psi_{m{\bf k}} \rangle $ is the Bloch state and $| g_n \rangle$ is the initial Wannier function (WF). For the low energy states, we have
\begin{equation}
|\Psi_{m{\bf k}} \rangle = \sum\limits_{\xi;\tau,l}\sum\limits_{\bf G} C_{n{\bf k}}^{(\xi;\tau,l)}({\bf G}) |\Psi_{{\bf k}_{\xi}^{l} + {\bf G}}^{(\xi;\tau,l)} \rangle.
\end{equation}
In this notation, $\xi \equiv \xi_{\pm} = \pm 1$ denotes the graphene valleys, $\tau = \tau_{\alpha}, \tau_{\beta}$ is graphene sublattice degree of freedom, $l$ stands for the layer index measured from the bottom to the top. ${\bf G} = n_1{\bf G}_1 + n_2 {\bf G}_2$ is the reciprocal lattice vector for TMLG (${\bf G}_{1,2}$ are the two components), and ${\bf k}_{\xi}^{l}$ is the TMLG valleys. Note that, $C_{n{\bf k}}^{(\xi;\tau,l)}({\bf G})$ can be calculated by diagonalizing the effective continuum model. On the other hand, the Bloch sum functions can be written explicitly as
\begin{equation}
|\Psi_{{\bf k}_{\xi}^{l} + {\bf G}}^{(\xi;\tau,l)} \rangle = \sum\limits_{{\bf L}, {\bf R}} e^{i({\bf k}_{\xi}^{l} + {\bf G}). \mathcal{D} [{\textrm {sign}}(d)\frac{\theta}{2}] ({\bf L} + {\bf R} + {\bf d})} |\tau_{\alpha},l, {\bf L} + {\bf R} + {\bf d}  \rangle.
\end{equation}
Here, ${\bf k}_{\xi}^{l} = {\bf k} + {\bf K}_{\xi}^{FLG} - {\bf K}_{\xi}^{l}$ is the wave vector of graphene, where ${\bf k}$ is the TMLG wave vector, and ${\bf K}_{\xi}^{FLG}$ is the graphene valley. ${\bf L}$ stands for the emerged moir\'e pattern, ${\bf R}$ represents the graphene lattice; ${\bf d} = dd_0\hat{z}$ denotes the layer stacking distance where $d_0$ is the distance between two graphene layers and $d$ is the layer index measured from the bottom to the top. $\mathcal{D} [{\textrm {sign}}(d)\frac{\theta}{2}]$ indicates a twist angle $\frac{\theta}{2}$ for the top partition and $-\frac{\theta}{2}$ for the bottom partition.

\medskip

The Hamiltonian matrix elements of the effective continuum model can be expressed as
\begin{align}
\nonumber
\langle \Psi_{{\bf k}_{\xi}^{l} + {\bf G}}^{(\xi;\tau_{\alpha},l)}| \hat{H}({\bf k}) |\Psi_{{\bf k}_{\xi}'^{l'} + {\bf G}'}^{(\xi;\tau_{\beta},l')} \rangle
= \langle \Psi_{{\bf k}_{\xi}^{l} + {\bf G}}^{(\xi;\tau_{\alpha},l)}| \hat{H}_0({\bf k}) |\Psi_{{\bf k}_{\xi}'^{l'} + {\bf G}'}^{(\xi;\tau_{\beta},l')} \rangle + \langle \Psi_{{\bf k}_{\xi}^{l} + {\bf G}}^{(\xi;\tau_{\alpha},l)}| \hat{H}_T |\Psi_{{\bf k}_{\xi}'^{l'} + {\bf G}'}^{(\xi;\tau_{\beta},l')} \rangle \\ \nonumber
=\delta_{\xi {\xi}'}\delta_{sign(l),sign(l')}\delta_{{\bf G}{\bf G}'} \times \sum\limits_{\bf R} e^{i({bf k}_{\xi}^{l}+{\bf G}).{\bf R}} \langle \tau_{\alpha},l,0+{\bf d} ||\tau_{\beta},l',{\bf R}+{\bf d}'  \rangle \\
+ \delta_{\xi {\xi}'}\delta_{l=\pm 1, l'=\mp 1} \times [T_1 \delta_{{\bf G},{\bf G}'} + T_2 \delta_{{\bf G},{\bf G}'+\xi{\bf G}_1} + T_3\delta_{{\bf G},{\bf G}'+\xi({\bf G}_1 + {\bf G}_2)}].
\end{align}
The matrices $T_{1,2,3}$ read as

\begin{align}
\nonumber
T_1 =
\begin{bmatrix} u & u' \\ u' & u \end{bmatrix},
\end{align}
\begin{align}
\nonumber
T_2 =
\begin{bmatrix} u & u'\omega^{-\xi} \\ u'\omega^{\xi} & u \end{bmatrix},
\end{align}
\begin{align}
\nonumber
T_3 =
\begin{bmatrix} u & u'\omega^{\xi} \\ u'\omega^{-\xi} & u \end{bmatrix},
\end{align}
in which, $\omega = e^{2\pi i/3}$; $u$ = 0.0797 eV and $u'$ = 0.0975 eV describe the relaxation effect.

\medskip
The WFs of a TMLG can be constructed from the $p_z$ orbitals together with an envelope function. If the twisted interlayer coupling is negligible, the low-energy Bloch states can be regarded as the folded
band structure of FLG.
The WFs can be expressed as
\begin{equation}
|g_n\rangle = \frac{1}{2}\sum\limits_{\xi} \sum\limits_{\tau, l, {\bf L},{\bf R}}
e^{i{\bf K}_\xi^{FLG} . {\bf r}} f_n^{(\xi, \tau, l)} ({\bf r}) |\tau, l, {\bf L} + {\bf R} + {\bf d}  \rangle.
\end{equation}
In this notation, $e^{i{\bf K}_{\xi} . {\bf r}}$ is the a high frequency factor, $|\tau, l, {\bf L} + {\bf R} + {\bf d}  \rangle$ is the $p_z$ orbital. $f_n^{(\xi, \tau, l)} ({\bf r})$ is the smooth envelope function in moir\'e length scale, in which $n$ stands for the index of the WFs. Note that, the system preserves the time-reversal symmetry for the two-valley TBM. We consider the real-value WFs by setting the limit $f_n^{(\xi_{+}, \tau, l)}$ = $f_n^{(\xi_{-}, \tau, l)}$ $\equiv$ $f_n^{(\tau, l)}$.
The initial condition of WFs can be established by using the constraint $f_1^{(\tau_{\alpha}, -1)} ({\bf r}) = G({\bf r} - {\bf r}_1^{\textrm {hex}})$, $f_2^{(\tau_{\beta}, -1)} ({\bf r}) = G({\bf r} - {\bf r}_2^{\textrm {hex}})$, $f_3^{(\tau_{\beta}, 1)} ({\bf r}) = - G({\bf r} - {\bf r}_1^{\textrm {hex}})$, and $f_4^{(\tau_{\alpha}, -1)} ({\bf r}) = - G({\bf r} - {\bf r}_2^{\textrm {hex}})$ with $G({\bf r} - {\bf r}_i^{\textrm {hex}})$ being the Gaussian function localized at the hexagonal site. These initial WFs satisfy the orbital character (flat bands) and lattice symmetry (hexagonal site symmetry, sublattice equivalence, and $C_{2x}$ symmetry).
In order to select the real-valued WFs, we set the limit $f_n^{(\xi_+;\tau,l)} = f_n^{(\xi_-;\tau,l)} = f_n^{(\tau,l)}$, for which the equivalence of the two valleys is established. Then,
\begin{equation}
|g_n\rangle = \frac{1}{2} \sum\limits_{\tau, l, {\bf L},{\bf R}}
cos({\bf K}_\xi^{FLG} . {\bf r}) f_n^{( \tau, l)} ({\bf r}) |\tau, l, {\bf L} + {\bf R} + {\bf d}  \rangle.
\end{equation}
\medskip

Therefore, the initial guess for the Bloch sum functions is now obtained by inserting the Bloch states from Eq. (9) and the WFs from Eq. (12) into Eq. (7). Now we take the singular value decomposition of $\psi^{(0)}_{n{\bf k}}$, it reads
\begin{equation}
| \psi^{(1)}_{n{\bf k}} \rangle = \sum\limits_{m} |\Psi_{m{\bf k}} \rangle (A_{\bf k}S^{-1/2}_{\bf k})_{mn},
\end{equation}
where
\begin{equation}
\nonumber
A_{\bf k} = U_{\bf k}\Sigma_{\bf k}V^{\dagger}_{\bf k},
\end{equation}
\begin{equation}
\nonumber
S^{-1/2}_{\bf k} = V_{\bf k} \frac{1}{\sqrt{\Sigma^{\dagger}_{\bf k}\Sigma_{\bf k}}} V^{\dagger}_{\bf k}.
\end{equation}
The Bloch sum functions can be obtained by projecting $\psi^{(1)}_{n{\bf k}}$ onto the subspace spanned by flat bands as done for the Bloch bands in ref. \cite{Bloch}. Particularly,
\begin{equation}
| \psi_{n{\bf k}} \rangle = \mathcal{P}^{f.b.}_{\bf k} | \psi^{(1)}_{n{\bf k}} \rangle
\end{equation}
with
\begin{equation}
\nonumber
\mathcal{P}^{f.b.}_{\bf k} = \sum\limits_{n} |\Psi_{n{\bf k}}^{f.b.} \rangle \langle |\Psi_{n{\bf k}}^{f.b.}|.
\end{equation}

\medskip
The number of ${\bf k}$ states is properly truncated so that it is sufficient to cover the main features of electronic property of TMLG. In fact, our effective TBM can reproduce well the band structures from the DFT method for arbitrary twist angles near the magic angle. Furthermore, the magnetic-field effect is considered by adding the Peierls phases to the effective Hamiltonian in the Bloch basis. In this way, the effective TBM enables to reduce significantly the computational cost, making the calculations of LLs at the magic angle practical.

\subsection{Kubo formula}

By assuming a weak applied DC electric field for electron transport, one can use a linear-response theory (or Kubo formula) to calculate conductivity.
Here, we employ the Kubo formalism in our studies of the QHE in graphene-based twisted systems. The expression for the DC quantum Hall conductivity (QHC) can be written as \cite{Kubo}

\begin{equation}
\sigma_{xy} = \frac {ie^2 \hbar} {{\cal S}}
\sum\limits_{\alpha}\,\sum\limits_{\beta \neq \alpha}\, \left(f_{\alpha} - f_{\beta}\right)\,\left[
\frac {\langle \alpha  |\dot{\mbox{\boldmath$u$}}_{x}| \beta\rangle  \langle \beta |\dot{\mbox{\boldmath$u$}}_{y}|\alpha \rangle} {(E_{\alpha}-E_{\beta})^2 + \Gamma ^2}\right] \ .
\label{eqn:5}
\end{equation}
In Eq.\,\eqref{eqn:5}, ${\cal S}$ stands for the area of field-extended unit cell, $E_{\alpha (\beta)}$ and $|\alpha (\beta) \rangle$ are, respectively, the eigen-values and eigen-functions of the Hamiltonian in Eq.\,\eqref{eqn:3}, $f_{\alpha (\beta)} = \{ 1 + \exp \{ (E_{\alpha (\beta)} - E_F)/k_B T  \}  \}^{-1}$ is the Fermi-Dirac distribution function at temperature $T$ with Fermi energy $E_F$, and $\Gamma$ ($\sim 1\,$meV) is the lifetime broadening of electrons. $\dot{\mbox{\boldmath$u$}}_{x(y)}$ is the velocity operator along the $x(y)$ direction. The velocity matrix element $\langle \alpha|\dot{\mbox{\boldmath$u$}}_{x(y)}| \beta \rangle$ can be computed using the gradient approximation of the forms \cite{gradient}

\begin{eqnarray}
\label{eqn:6}
\langle \alpha  |\dot{\mbox{\boldmath$u$}}_{x}| \beta\rangle &\approx& \frac{1}{\hbar } \left \langle \alpha \left|\frac{\partial H}{\partial k_x}\right|\beta\right\rangle\ ,\\
\label{eqn:7}
\langle \alpha  |\dot{\mbox{\boldmath$u$}}_{y}| \beta\rangle &\approx& \frac{1}{\hbar } \left\langle \alpha \left|\frac{\partial H}{\partial k_y}\right|\beta \right\rangle\ .
\end{eqnarray}
It is noted from Eq.\,\eqref{eqn:5} that a system has a finite QHC as both factors in Eqs.\,\eqref{eqn:6} and \eqref{eqn:7} are non-vanishing. This condition can only be satisfied if the wave functions of the initial and final states in transition acquire the same oscillation mode. Such a selection rule for excitation will be extensively discussed for each system introduced in Section\,\ref{sec3}.

\section{Results and Discussion}
\label{sec3}

\subsection{Monolayer graphene and AB bilayer graphene}

\begin{figure}[htbp]
\begin{center}
\includegraphics[width=0.6\linewidth]{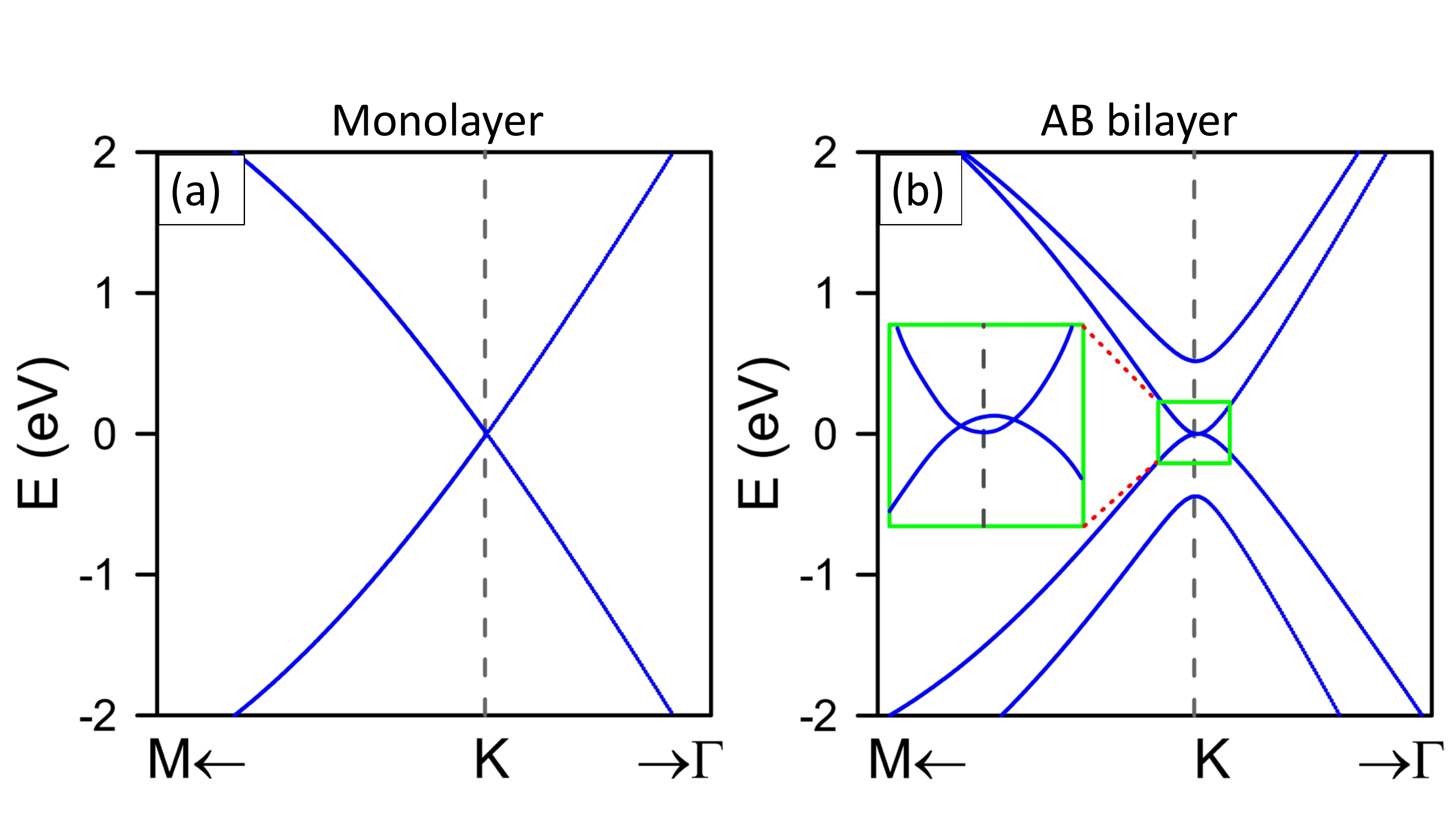}
\end{center}
\caption{(color online) Band structures of (a) monolayer graphene and (b) AB bilayer graphene along the high symmetry  points $M, K$ and $\Gamma$.}
\label{Fig3}
\end{figure}

The band structure of a monolayer graphene consists of six Dirac cones at the corner points ($K$ and $K^\prime$) of the hexagonal first Brillouin zone. The valence and conduction bands meet at the Dirac points, making monolayer graphene a gapless semiconductor. Both the upward and downward cones are isotropic in $\mbox{\boldmath$k$}$ space and they are mostly symmetric in the vicinity of Fermi energy $(E_F = 0)$. However, these cones become anisotropic at sufficiently high energies, as seen in Fig.\,\ref{Fig3}(a). On the other hand, AB bilayer graphene possesses two pairs of asymmetric valence and conduction bands, as displayed in Fig.\,\ref{Fig3}(b). Such an asymmetry in the bands results from the breaking down of lattice-mirror symmetry in the Bernal stacking. Moreover, the pair of subbands closer to $E_F$ = 0 are overlapped slightly, as shown in the insert of Fig.\,\ref{Fig3}(b). For both monolayer graphene and AB bilayer graphene, our TBM prediction for low-lying bands is in good agreement with those calculated using DFT method\,\cite{dftmono1, dftab1} and experimental measurements\,\cite{expmono1, expmono2, expab1} as well.
\medskip

\begin{figure}[htbp]
\begin{center}
\includegraphics[width=0.6\linewidth]{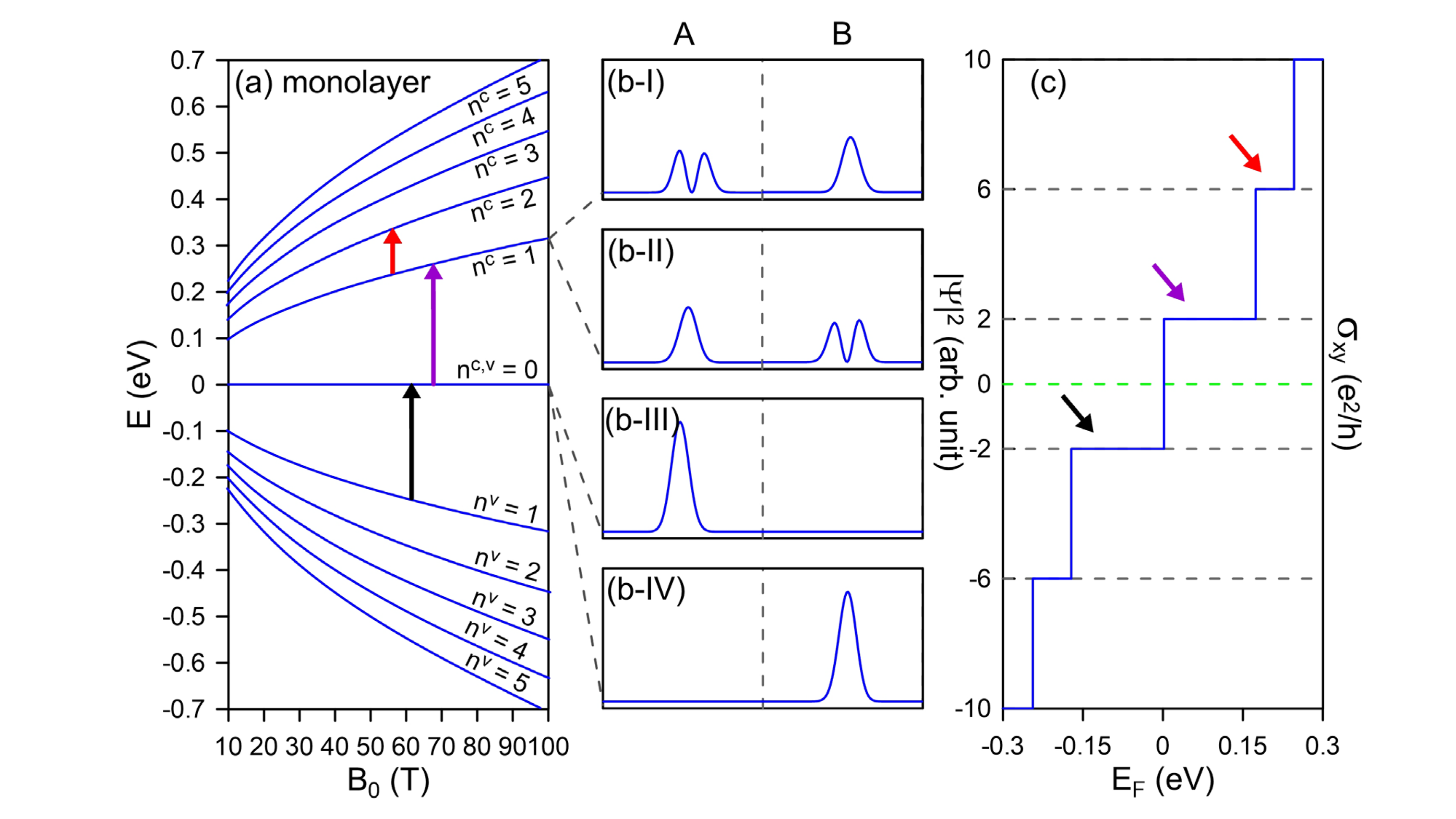}
\end{center}
\caption{(color online) (a) Magnetic-field-dependent LL energies of monolayer graphene. The notations $n^{c,v}\ ( = 0, 1, 2,\cdots)$ stand for the conduction and valence LLs, respectively, where $n$ is the index of LLs relating to the number of zero nodes in corresponding wave functions. The vertical arrows indicate the inter-LL transitions. (b-I) through (b-IV) present the magnitudes of wave functions with respect to $n^{c,v} = 0$ and $n^{c} = 1$. (c) Fermi-energy-dependent QHC at $B_0= 30\,$T. The arrows in (c) show discrete plateaus in correspondence with vertical transitions between LLs  in (a).}
\label{Fig4}
\end{figure}

The LLs of MLG exhibit a unique field-dependent spectrum, as found from Fig.\,\ref{Fig4}(a). The zeroth LL, which is flat and non-dispersive at zero energy, is independent of magnetic field strength $B_0$. At higher and deeper energies, LLs resulting from the linear bands present a proportional relationship with $B_0$, i.e., $E_n \propto \sqrt{nB_0}$. Regarding the gap between two nearest-neighbor LLs, it decreases at higher or deeper energies, which is associated with the nature of Dirac-cone band structure. With increasing $B_0$, this gap is enlarged gradually. Figures\,\ref{Fig4}(b-I) through \ref{Fig4}(b-IV) show the probability distribution (square of wave function, $|\Psi|^2$) of the $A$ and $B$ atoms for the $n^{c,v}$ = 0 and $n^{c}$ = 1 LLs. Here, each LL is four-fold degenerate due to the lattice symmetry and spin interaction\,\cite{degenerate}. For the degenerate LLs, the probability distribution is identical for the spin-up and spin-down states. On the other hand, the equivalence of $A$ and $B$ sublattices gives rise to two different configurations for $|\Psi|^2$, e.g., one of them is dominated by $A$ atom [see Figs.\,\ref{Fig4}(b-I) and \ref{Fig4}(b-III)] while the other is governed by $B$ atom [Figs.\,\ref{Fig4}(b-II) and \ref{Fig4}(b-IV)]. The index of each LL, which is determined from the number of zero-modes for the dominant sublattice, plays an important role in revealing a selection rule for the inter-LL transitions in transport and optical properties.
\medskip

The $E_F$-dependent QHC of MLG displays a unique step structure, as presented in Fig.\,\ref{Fig4}(c), following the sequence of $4(m-1/2)e^2/h$ in which $m$ is an integer. Interestingly, the  $\sigma_{xy}$ = 0 plateau is missing due to the quasi-particle excitation in graphene. Specifically, the $n^{c,v}$ = 0 LL exhibits both electron-like and hole-like characteristics, leading to unusual half-integer QHE in graphene. The arrows in Fig.\,\ref{Fig4}(c) show the discrete plateaus resulting from the corresponding vertical transitions between LLs indicated in Fig.\,\ref{Fig4}(a). For example, the plateau for $\sigma_{xy} = 2\,e^2/h$ is connected to the excitation from $n^{c,v} = 0$ to $n^c = 1$ LLs. We further observe that the $n^c = 1$ LL consists of both the $n = 0$ and $n = 1$ modes of probability distribution, as seen in Figs.\,\ref{Fig4}(b-I) and \ref{Fig4}(b-II). Therefore, the $n^{c,v}$ = 0 $\rightarrow$ $n^c = 1$ transition produces a nonzero value for the velocity matrix element terms in Eq.\,\eqref{eqn:5}, which yields a finite QHC. In general, the selection rule for excitation is defined as $\Delta n = |n_f - n_i| = 1$, where $n_i$ and $n_f$ stand for the LL indices with respect to the initial and final states, respectively. Moreover, the size of QHC plateaus becomes smaller for larger and deeper $E_F$, in  agreement with alteration of LL energy spacing. The half-integer QHE in graphene has been suggested by different theory groups\,\cite{qhemono1, qhemono2, qhemono3} and verified by experimental measurements\,\cite{qhemono4, qhemono5}. The quantization obtained by our computation method is in accord with previous theoretical and experimental studies for monolayer graphene, indicating the suitability of our proposed model theory in calculating the QHE of this system. Most importantly, our theoretical method enables a thorough explanation for the occurrence of unique QHC step structures.
\medskip

\begin{figure}[htbp]
\begin{center}
\includegraphics[width=0.6\linewidth]{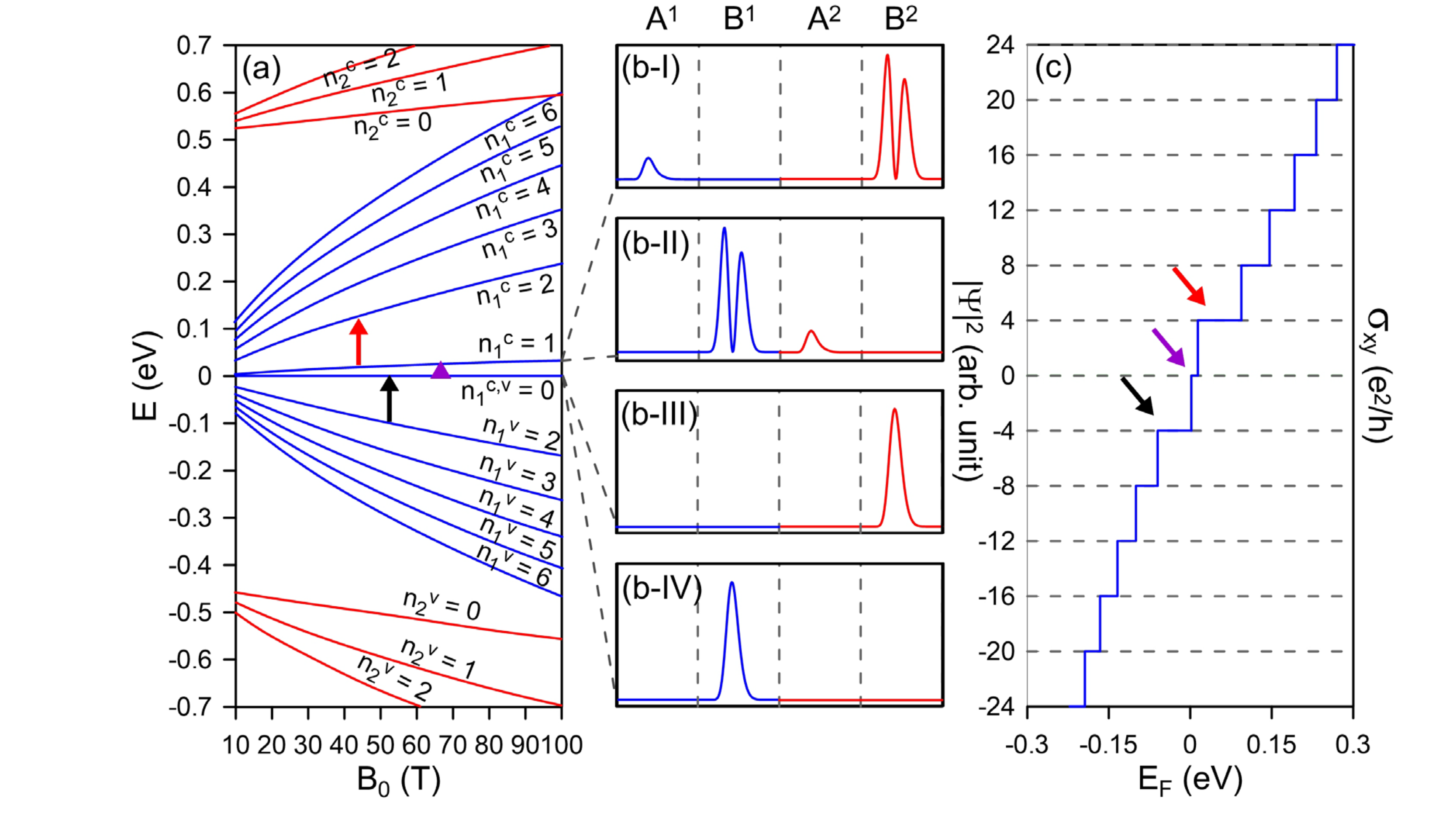}
\end{center}
\caption{(color online) (a) $B_0$-dependent LLs of AB bilayer graphene. The vertical arrows indicate the inter-LL transitions.  Panels (b-I)  through (b-IV) present the wave functions squared for $n^{c,v} = 0$ and $n^{c} = 1$. (c) $E_F$-dependent QHC at $B_0 = 30\,$T. The arrows in (c) show discrete plateaus coming from vertical transitions between LLs in (a).}
\label{Fig5}
\end{figure}

For AB BLG, the $B_0$-dependent LL spectrum can be classified into two distinct groups, appearing as blue ($n_1^{c,v}$) and red ($n_2^{c,v}$) lines in Fig.\,\ref{Fig5}(a), corresponding to two pairs of conduction and valence energy bands. Both LL groups, excluding the $n_1^{c,v} = 0$ at zero energy, possess a linear dependence on magnetic field strength $B_0$. The physical properties at low energies are largely dominated by the $n_1^{c,v}$ group. Similar to MLG, LLs are four-fold degenerate here. Of special interest is the slight overlap of low-lying pair of energy bands, which results in peculiar LLs near zero energy, e.g., spectrum asymmetry and small splitting of the $n_1^{c,v} = 0$ and $n_1^{c} = 1$ LLs. In fact, these two LLs overlap as the magnetic field is sufficiently weak ($B_0 \leq 10\,$T), but split otherwise. Their probability distributions on $A$ and $B$ sublattices of the first ($A^1$ and $B^1$) and second ($A^2$ and $B^2$) graphene layers are displayed in Fig.\,\ref{Fig5}(b). Interestingly, the probability distribution for each LL is strongly dominated by either $B^1$ or $B^2$ sublattice. The minor role of $A$ atoms might be accounted for by suppression of spin-orbital-coupling for $A^1$ and $A^2$ sublattices due to their identical $(x,y)$-projection.
\medskip

AB BLG has a unique $E_F$-dependent QHC spectra,  whereby the plateaus appear at $4me^2/h$, as clearly demonstrated in Fig.\,\ref{Fig5}(c). We notice that the $n_1^{c} = 1$ LL contains both $n = 0,\,1$ oscillation modes although the $n = 0$ mode exhibits a relatively small amplitude. This facilitates the excitation between $n_1^{c,v} = 0$ and $n_1^{c} = 1$ LLs, giving rise to a QHC plateau at zero energy, as indicated by purple arrows in Figs.\,\ref{Fig5}(a) and \ref{Fig5}(c). The AB BLG goes along with the selection rule $\Delta n = 1$ for transitions, similar to that of MLG. The narrow QHC plateau, corresponding to $n_1^{c,v} = 0\rightarrow n_1^{c} = 1$ transition, is correlated with the energy difference between these two LLs, and therefore it will vanish for very small $B_0$. It is worth mentioning that the energy splitting between the $n_1^{c,v} = 0$ and $n_1^{c} = 1$ LLs is infinitesimal for laboratory magnetic field so that it could not be resolved by transport measurement. Our numerical calculations show that AB BLG exhibits a special QHC step of 8$e^2/h$ at a suitable field strength when both $n_1^{c,v} = 0$ and $n_1^{c} = 1$ LLs are filled. This signifies the consistency between $B_0$-dependent LL spectrum and QHC. Our predicted filling factors of $\pm 4,\,\pm 8,\,\cdots$ are found identical to those in previous theoretical\,\cite{qhemono3, qheab} and experimental\,\cite{qhemono4} verification.

\subsection{Twisted bilayer graphene and Twisted double bilayer graphene}

\subsubsection{Twisted bilayer graphene}

\begin{figure}[htbp]
\begin{center}
\includegraphics[width=0.6\linewidth]{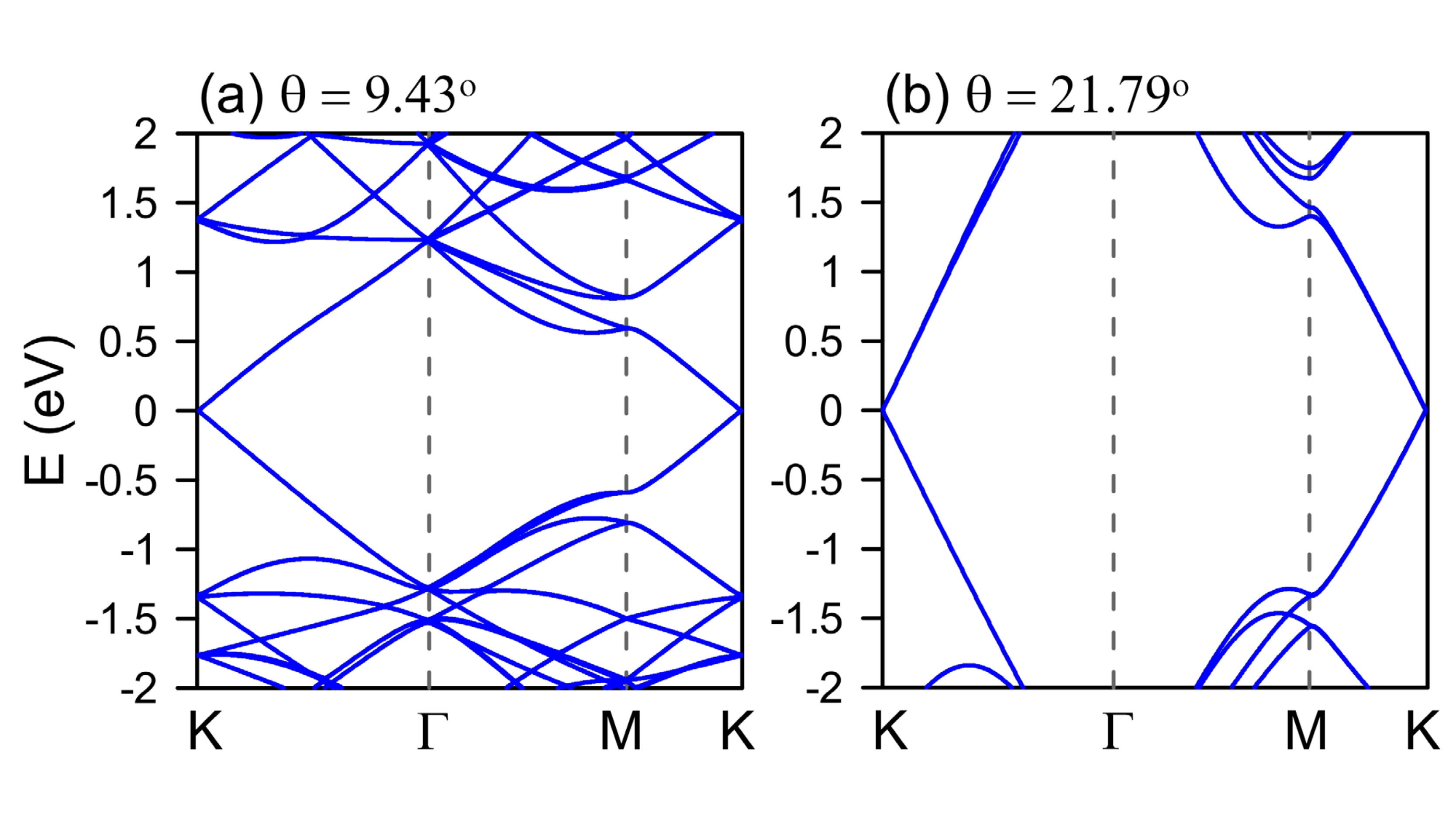}
\end{center}
\caption{(color online) Band structures of TBLG with a twist angle of (a) $\theta = 9.43^{\rm o}$ and (b) $\theta = 21.79^{\rm o}$.}
\label{Fig6}
\end{figure}

The band structures of $\theta = 9.43^{\rm o}$ and $\theta = 21.79^{\rm o}$ TBLG present the Dirac-cone shape at the $K$ point, as shown in Figs.\ref{Fig6}(a) and \ref{Fig6}(b). In the vicinity of zero energy, the linear dispersion resembles that of MLG. However, the conduction and valence bands are eight-fold degenerate, which is twice the band degeneracy of MLG. This characteristic plays a distinctive role in QHE. At higher energies, there exists a saddle point $M$ as well as a hole pocket at $\Gamma$ point. However, the main features of band structure are similar for various twist angles, but the energy scales are quite different. We find that within the same energy range, the $\theta = 9.43^{\rm o}$ hetero-structure shows more subbands compared with the case of $\theta = 21.79^{\rm o}$. That is, the energy scale is retracted for decreasing twist angle. The robust dependence of electronic properties on a twist angle can be explained by the band-folding effect. Specifically, the relative rotation of a graphene layer leads to the extension of unit cell, causing band folding in momentum space. Consequently, the first Brillouin zone is folded with a greatly reduced size. Our calculations reveal that the retraction of energy scale and the reduction of folded Brillouin zone have a proportional relationship.
\medskip

\begin{figure}[htbp]
\begin{center}
\includegraphics[width=0.6\linewidth]{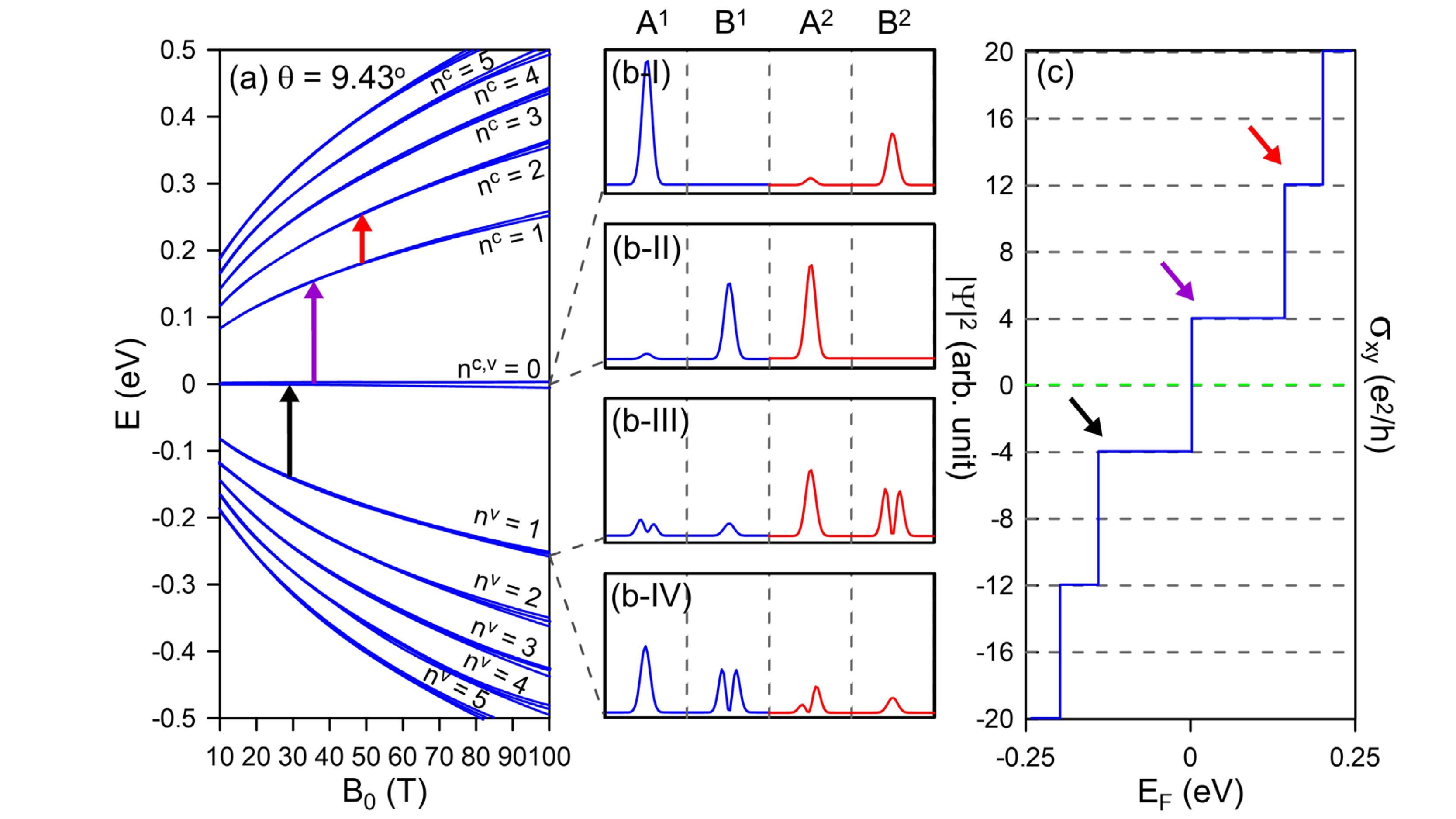}
\end{center}
\caption{(color online) (a)  The magnetic-field-dependent Landau energies of twisted bilayer graphene with a twist angle $\theta = 9.43^o$. The vertical arrows indicate the inter-LL transitions. Panels (b-I) through (b-IV) present the square of the wave functions for $n^{c,v}$ = 0 and $n^{v}$ = 1.  Panel (c) shows the Fermi-energy-dependent quantum Hall conductivity at $B_0$ = 30 T. The arrows in (c) show the discrete plateaus coming from the corresponding vertical transitions between the LLs indicated in (a).  }
\label{Fig7}
\end{figure}

\begin{figure}[htbp]
\begin{center}
\includegraphics[width=0.6\linewidth]{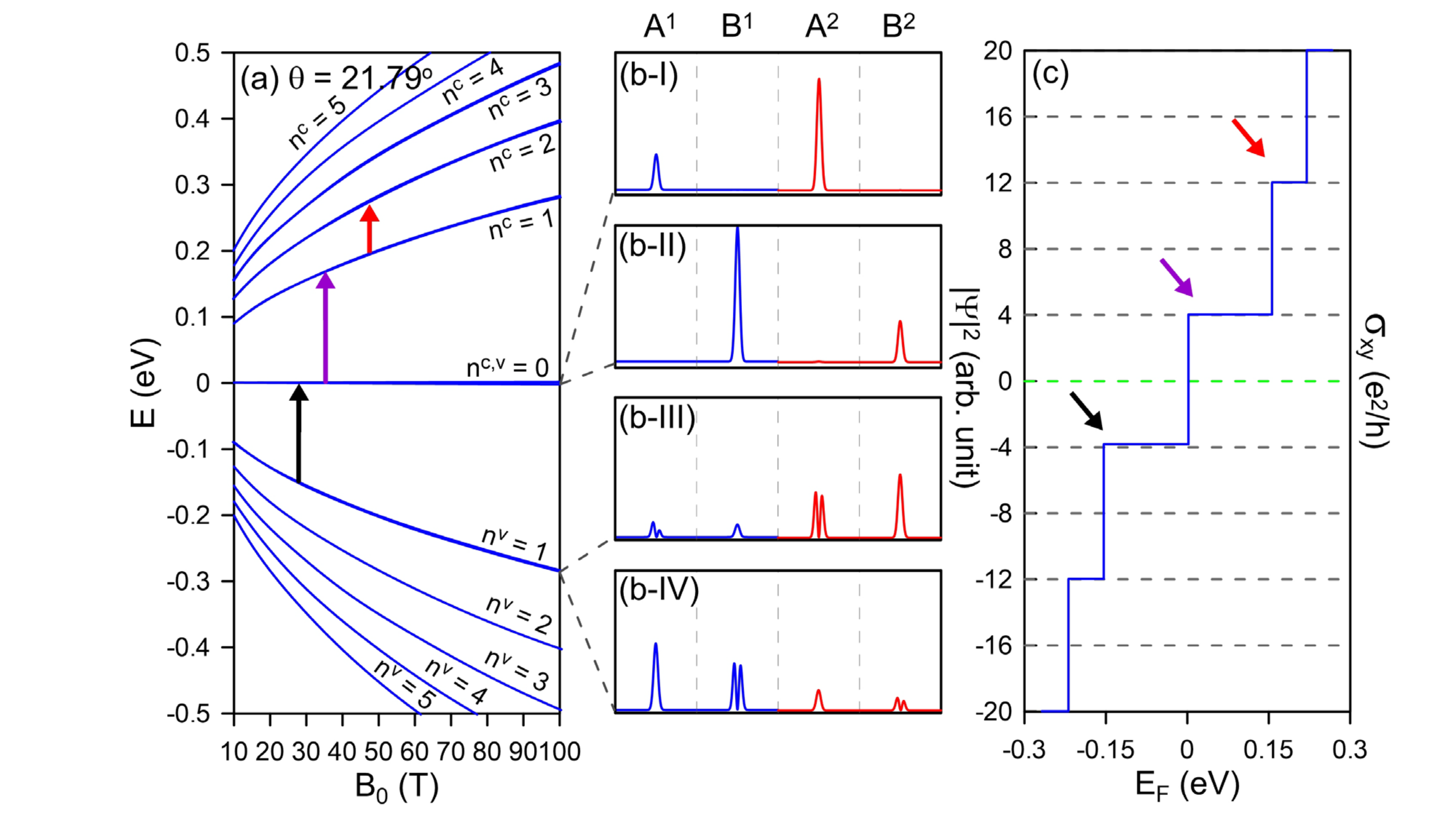}
\end{center}
\caption{(color online) (a) $B_0$-dependent LL energies of twisted bilayer graphene with a twist angle $\theta = 21.79^{\rm o}$. The vertical arrows indicate the inter-LL transitions. Panels (b-I) through (b-IV) display the wave function amplitudes of $n^{c,v} = 0$ and $n^{v} = 1$. Panel (c) presents the $E_F$-dependent QHC at $B_0 = 30\,$T. The arrows in (c) show discrete plateaus,  corresponding to arrows in (a) for vertical transitions between the LLs.}
\label{Fig8}
\end{figure}

The $B_0$-dependence of LLs in TBLG is found similar to that of MLG because their low-lying bands can be viewed as two overlapped Dirac cones. Figures\,\ref{Fig7}(a) and \ref{Fig8}(a) show the LL spectra of $\theta = 9.43^{\rm o}$ and $\theta = 21.79^{\rm o}$ TBLG, respectively. Here, the nearly-flat zeroth LLs, as well as the square root dependence of LL energies on $n>0$ and $B$, still remain at low $B_0$ fields. This implies that the atomic couplings between two layers are extremely weak at low energies. As a matter of fact, two LL groups $n_1^{c,v}$ and $n_2^{c,v}$ overlap each other. This explains well why each LL is eight-fold degenerate, which is twice that of MLG. This characteristic is also different from four-fold degenerate LLs of AB BLG for which two LL groups are well separated. Nevertheless, a sufficiently high $B_0$ can destroy the degeneracy of LLs for which the splitting of degenerate LLs is manifest. The critical field at which such LL splitting occurs increases with twist angle $\theta$.
\medskip

The probability distribution provides important clues for inter-LL transitions. Figures\,\ref{Fig7}(b-I) through \ref{Fig7}(b-IV) present the probability distributions for $n^{c,v} = 0$ and $n^{v} = 1$ LLs with $\theta = 9.43^{\rm o}$ TBLG. Here, the most influential sublattices of each LL can be either $A^1$ and $B^2$ or $A^2$ and $B^1$. For $\theta = 21.79^{\rm o}$, on the other hand, the dominant sublattices are $A^{1,2}$ or $B^{1,2}$, as seen in Figs.\,\ref{Fig8}(b-I) through \ref{Fig8}(b-IV). This difference is associated with the change of relative positions of atoms under the rotating procedure. In both twisted systems, two graphene layers play the equivalent role in the LL probability distributions. This is in contrast with AB BLG where only $B$ atoms have a crucial contribution to the $|\Psi|^2$ of LLs. It is noted that, the existence of zero mode in the $n^{v} = 1$ LL state allows for the excitation between $n^{c,v} = 0$ and $n^{v} = 1$ LLs, leading to unique QHC plateaus. The selection rule for excitation in TBLG and MLG, as well as in AB BLG, are equivalent, given by $\Delta n = 1$.
\medskip

The $E_F$-dependent QHC for $\theta = 9.43^{\rm o}$ and $21.79^{\rm o}$ TBLG exhibit the sequence of $8(m -1/2)\,e^2/h$ (Figs.\,\ref{Fig7}(c) and \ref{Fig8}(c)), corresponding to the eight-fold degenerate LLs at $B_0 = 30\,$T. Our theoretical results are in good agreement with the previous studies of QHC in TBLG\,\cite{PRB2012Moon, PRB2012Fal, PRL2011Lee, PRL2012San} for various twist angles $\theta$. The size of plateaus becomes wider for larger $\theta$, which correctly reflects the LL energy spacing. At zero energy, QHC varies from $-4\,e^2/h$ to $4\,e^2/h$, in correspondence with the vertical transition from $n^{c,v} = 0$ to $n^c = 1$ as indicated by purple arrows. That is, there is no zero QHC step, which is different from a narrow zero plateau in AB BLG. Such a feature is consistent with $B_0$-induced LL splitting and its dependence on $\theta$ as discussed above. Consequently, the QHCs for $\theta = 9.43^{\rm o}$ and $21.79^{\rm o}$ are expected to be dissimilar at higher $B_0$. We emphasize that the QHCs of TBLG and MLG are quite similar because they both share the Dirac-cone band structures.

\subsubsection{Twisted double bilayer graphene}

\begin{figure}[htbp]
\begin{center}
\includegraphics[width=0.6\linewidth]{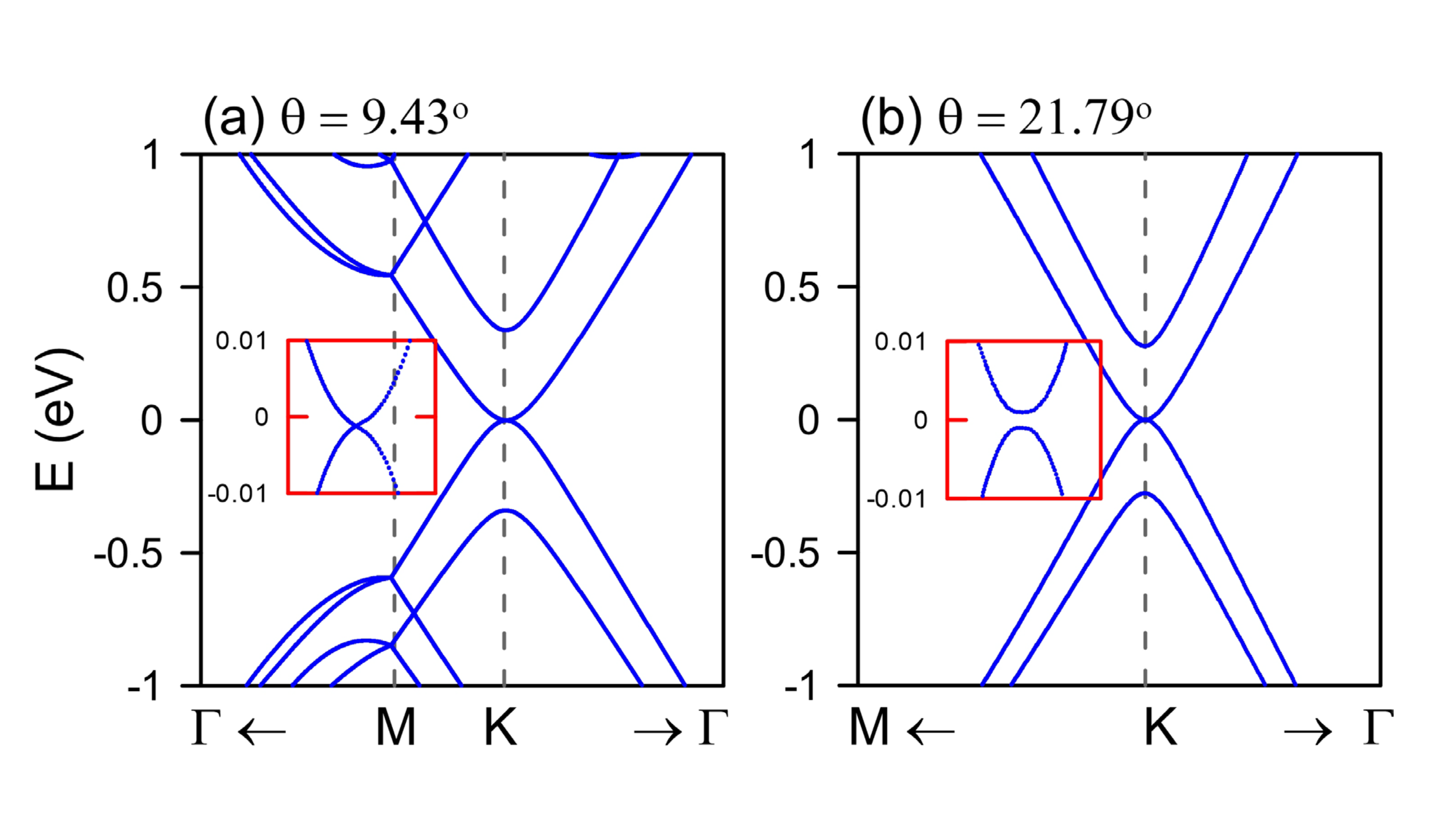}
\end{center}
\caption{(color online) Band structures of AB-AB TDBLG with twist angles of (a) $\theta = 9.43^{\rm o}$ and (b) $\theta = 21.79^{\rm o}$.}
\label{Fig9}
\end{figure}

The low-lying band structures of TDBLG display two pairs of parabolic conduction and valence bands, as in Figs.\,\ref{Fig9}(a) and \ref{Fig9}(b) for $\theta = 9.43^{\rm o}$ and $21.79^{\rm o}$, respectively. Here, each band is only doubly degenerate due to finite number of graphene layers in the system. Note that the couplings between two middle graphene layers are weak. Therefore, TDBLG can be regarded roughly as two Bernal BLG. For higher and deeper energy ranges, the degenerate bands split in the direction of $M\to\Gamma$. Such band splitting occurs at lower energies for smaller $\theta$. The zoom-in view of the band structure for $\theta = 9.43^{\rm o}$ in the inset of Fig.\,\ref{Fig9}(a) demonstrates that this system is a gapless semiconductor. For $\theta = 21.79^{\rm o}$, on the other hand, there is a narrow band gap in the vicinity of zero energy, as illustrated in the inset of Fig.\,\ref{Fig9}(b). Our calculated gap size is consistent with previous prediction\,\cite{double1, double2}. Similar to TBLG, the retraction of energy scale is proportional to the reduction of folded Brillouin zone as well as the twist angle.
\medskip

\begin{figure}[htbp]
\begin{center}
\includegraphics[width=0.6\linewidth]{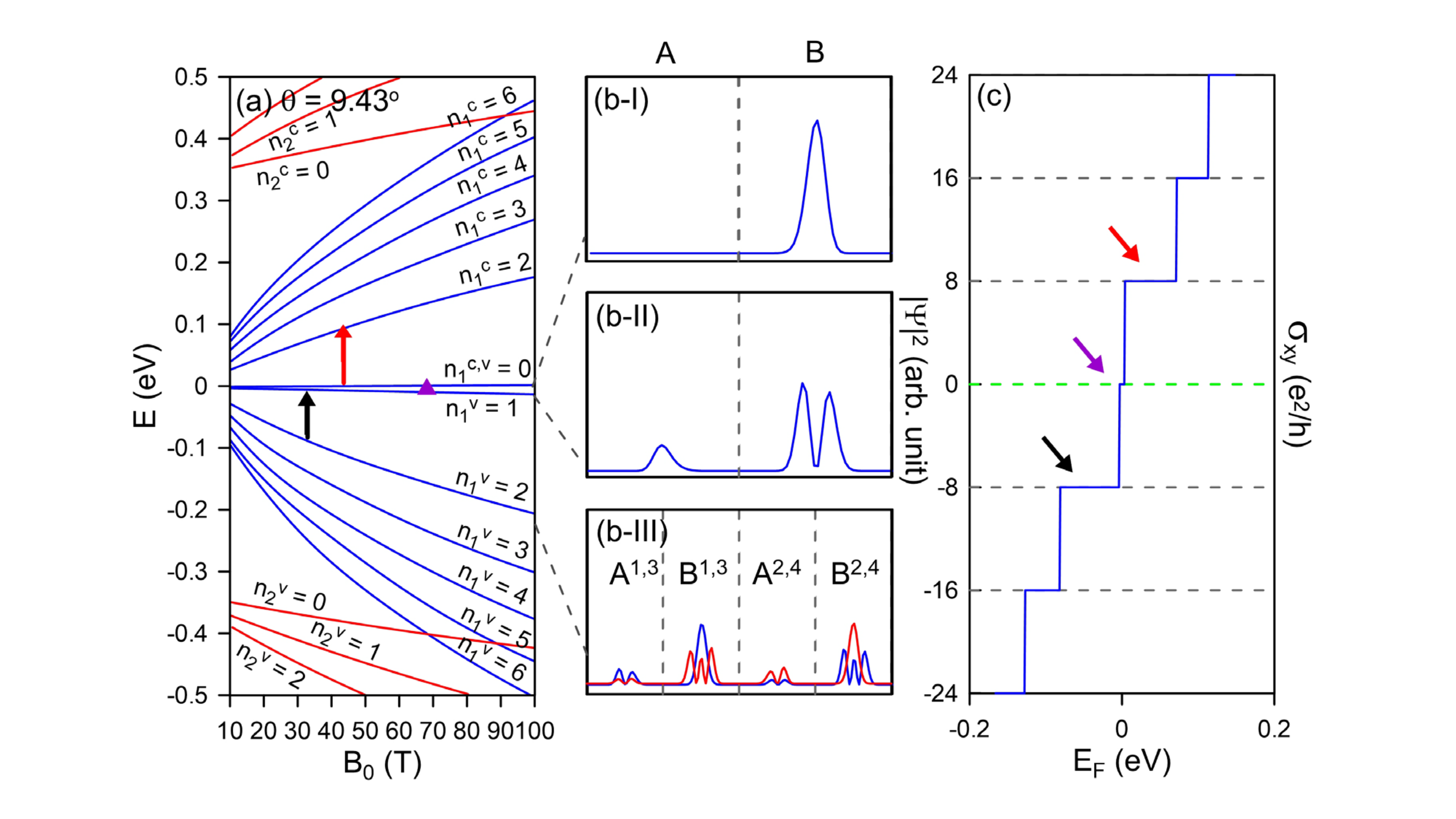}
\end{center}
\caption{(color online) (a) $B_0$-dependent Landau energies of AB-AB TDBLG with a twist angle $\theta = 9.43^{\rm o}$. The vertical arrows indicate the inter-LL transitions. Here, (b-I) through (b-III) present the corresponding wave function amplitudes of $n_1^{c,v} = 0$, $n_1^{v} = 1$ \& $2$ at $B_0 = 30\,T$. The $E_F$-dependent QHC is illustrated in (c). The arrows in (c) point to discrete plateaus, corresponding to vertical transitions between LLs indicated in (a).}
\label{Fig10}
\end{figure}

\begin{figure}[ht]
\begin{center}
\includegraphics[width=0.6\linewidth]{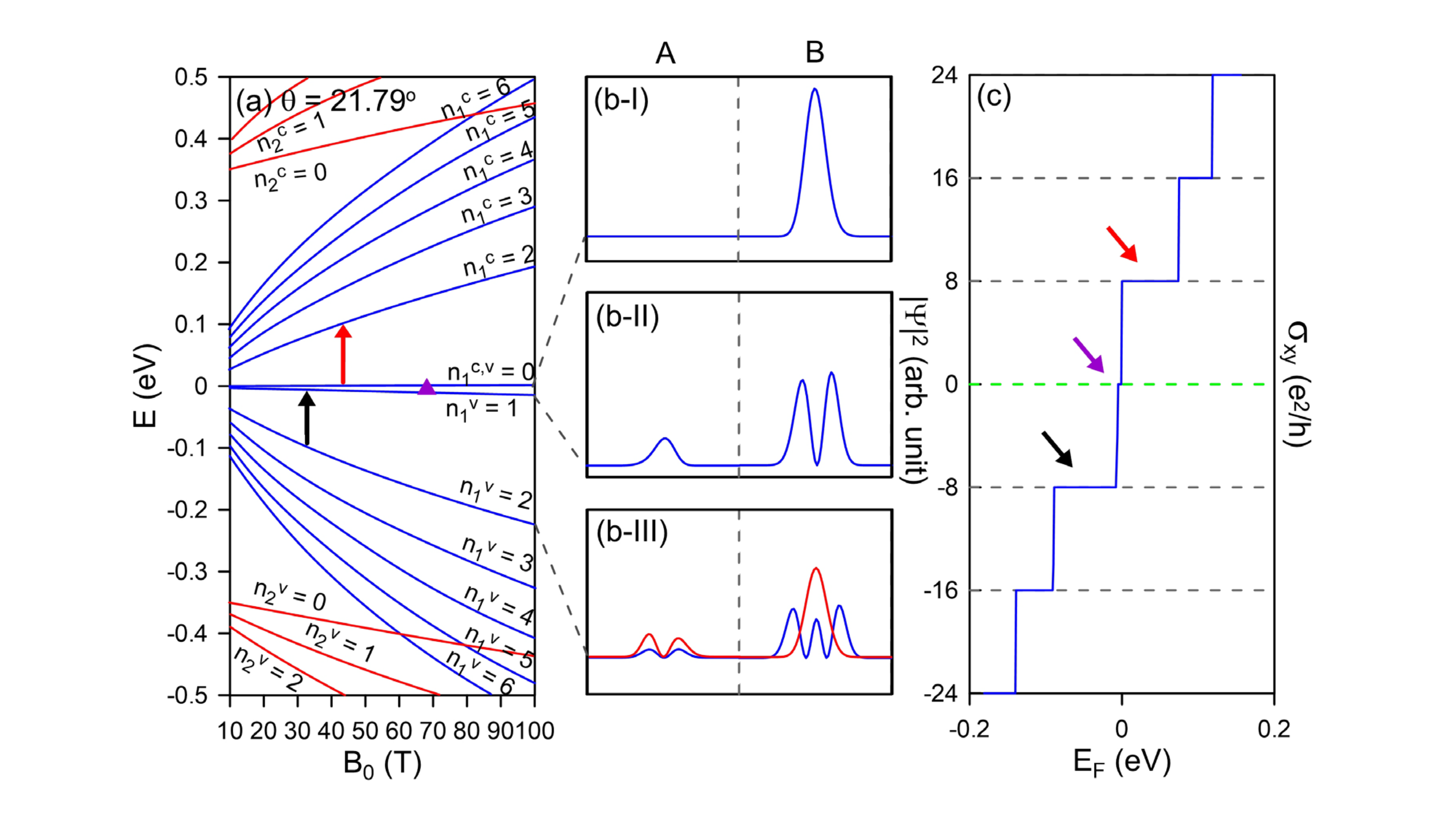}
\end{center}
\caption{(color online) (a) $B_0$-dependent LL energies of AB-AB TDBLG with $\theta = 21.79^{\rm o}$. The vertical arrows indicate the inter-LL transitions.  Panels (b-I) through (b-IV)  present the amplitudes of wave function corresponding to $n_1^{c,v} = 0$, $n_1^{v} = 1$ \& $2$ at $B_0 = 30\,$T. The $E_F$-dependent QHC is given in (c). The arrows in (c) indicate the discrete plateaus due to  vertical transitions between LLs indicated in (a).}
\label{Fig11}
\end{figure}

The $B_0$-dependent LL energy spectra of AB-AB TDBLG, shown in Figs.\,\ref{Fig10}(a) and \ref{Fig11}(a), can be classified into two groups, namely, $n_1^{c,v}$ and $n_2^{c,v}$. The crucial physical properties of these systems are mainly dominated by the $n_1^{c,v}$ LL group at low energies. There exists an overlap of the $n_1^{c,v} = 0$ and $n_1^{v} = 1$ LLs at low $B_0$, similar to that of AB BLG. This agrees with the equivalence of their energy dispersions as discussed above. In general, the main features of LLs resemble those of AB BLG since the interlayer interactions within each BLG component are much stronger than those between them. Despite the similarity in the $B_0$-dependence of LLs, the LL degeneracies of the two systems are distinctive due to the difference in number of graphene layers. Moreover, the probability distributions of TDBLG are more diversified as we will discuss next.
\medskip

Figures\,\ref{Fig10}(b) and \ref{Fig11}(b) present distributions of $|\Psi|^2$ for $\theta = 9.43^{\rm o}$ and $21.79^o{\rm o}$ TDBLG, respectively. In order to highlight twisting effect in the system, we combine the probability distributions of $A$ and $B$ atoms for each component AB BLG. For each LL, the combined $|\Psi|^2$ of $A$ sublattices on the upper and lower BLG are identical, and also for the $B$ sublattices. The LL indicies are determined from the number of zero modes in $|\Psi|^2$ on the dominant $B$ atoms. Surprisingly, we find the $n = 0$ \& $1$ modes for $n_1^{v} = 2$ LL, as seen in Figs.\,\ref{Fig10}(b-III) and \ref{Fig11}(b-III) by separated blue and red curves. This gives rise to a new excitation selection rule of $\Delta n = 2$ in addition to a conventional one $\Delta n = 1$. In particular, the transitions between $n_1^{c,v} = 0$ and $n_1^{c,v} = 2$ occur beside those between $n_1^{c,v} = 1$ and $n_1^{c,v} = 2$ as in three systems discussed above. The unusual characteristics of LLs are expected resulting from the special arrangement in the TDBLG lattice structure. Moreover, we find that the TDBLG is less sensitive to the twist angle $\theta$ compared with the TBLG because their electronic characteristics are dominated by the individual AB BLG components for large $\theta$.
\medskip

The $E_F$-dependent QHC of TDBLG with large $\theta$ follows the sequence $8\,me^2/h$ due to its eight-fold degenerate LLs. The step structures for $\theta = 9.43^{\rm o}$ and $21.79^{\rm o}$ TDBLG at $B_0 = 30\,$T are given in Figs.\ref{Fig10}(c) and \ref{Fig11}(c), respectively. Here, the narrow zero plateau like that of component AB BLG is also observed, which comes from the $n_1^{v} = 1\to n_1^{c,v} = 0$ LL transition. The size of this plateau is determined by the energy spacing between these two LLs, and is proportional to $B_0$. At sufficiently low $B_0$, the zero QHC will disappear so that the system will achieve the double QHC step of $16\,e^2/h$. Such a large QHC step is unique for the AB-AB TDBLG, which has not been reported for other graphene-related materials. Our prediction of the filling factors can be examined by  transport measurements, as done for MLG and AB BLG.

\subsubsection{Magic angle}

\begin{figure}[htbp]
\begin{center}
\includegraphics[width=0.9\linewidth]{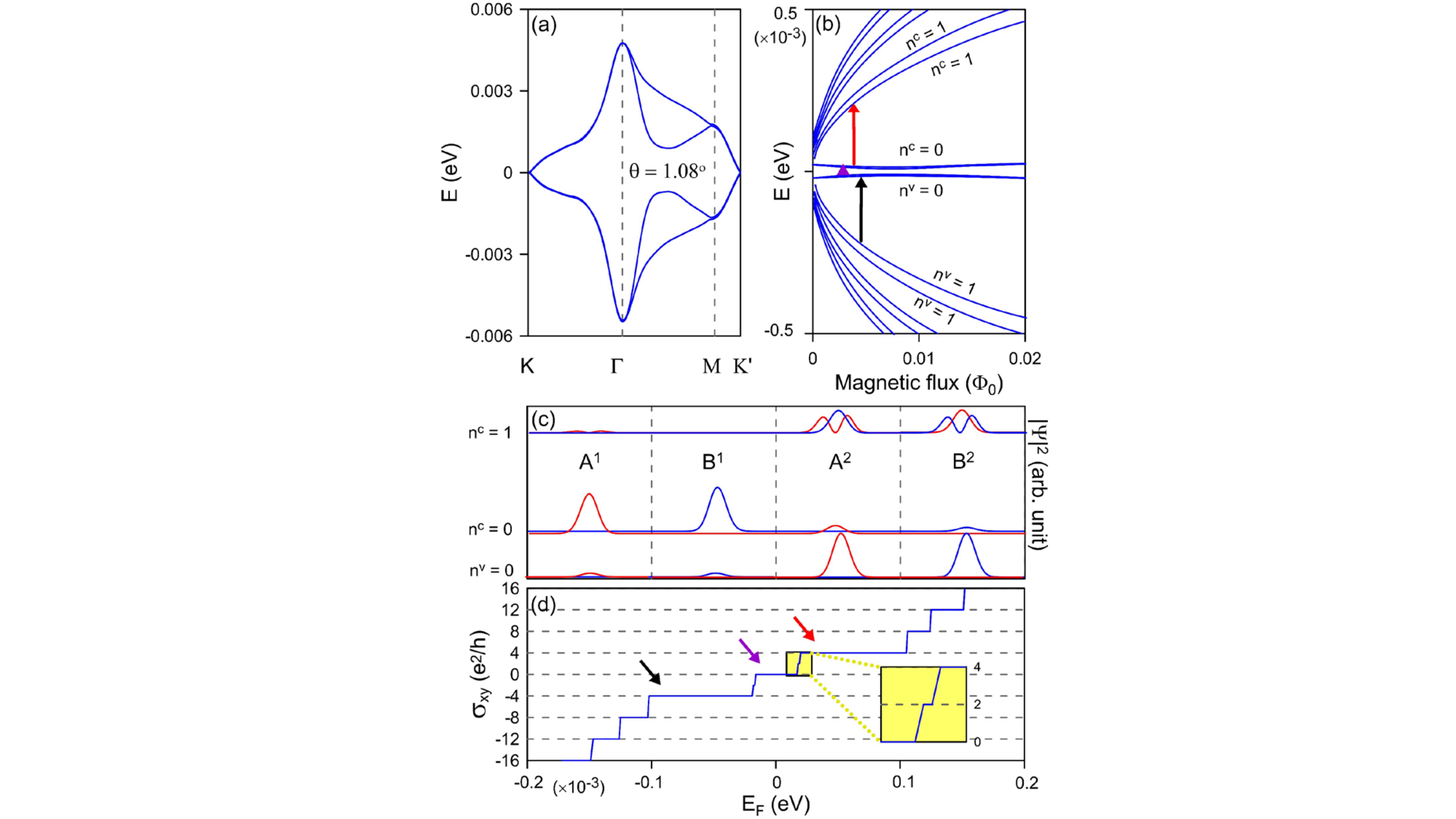}
\end{center}
\caption{(color online) (a) Band structure of TBLG at the magic angle $\theta = 1.08^{\rm o}$. (b) $\Psi$-dependent LLs at low energies. The vertical arrows indicate the inter-LL transitions. (c) Wave-function amplitudes for $n^{c,v} = 0$ and $n^{c} = 1$ at $B_0= 0.001\phi_0$, and (d) $E_F$-dependent QHC. The arrows in (d) show the discrete plateaus coming from the vertical transitions between LLs as indicated in (a).}
\label{Fig12}
\end{figure}

The band structure of TBLG at the magic angle $\theta = 1.08^{\rm o}$, shown in Fig.\ref{Fig12}(a), is calculated by using four-band TBM. This method has been demonstrated before as being suitable for generic twisted multilayer graphene with small twist angle\,\cite{effective}. The four low-lying bands present a linear dispersion near the $K$ and $K^\prime$ points as well as the saddle points near the $M$ point. These two pairs of valence and conduction energy bands are nearly overlapping around the $K$ and $K^\prime$ points while they split along $\Gamma M$. It is important to notice that the $K$ and $K^\prime$ points here are no longer equivalent like in the case of MLG and AB BLG. This is because the small $\theta$ greatly changes the crystal symmetry and consequently the electronic properties of the system.
\medskip

The $\Phi$-dependent LL energy spectrum is quite different from that of MLG with more complicated features. This results from significant couplings between two graphene layers, in contrast with the TBLG with larger twist angle. The two LLs in the vicinity of zero energy are separated by a small gap for a wide range of $\Phi$. Under an external magnetic field, the LL degeneracy is lifted and reduced from eight-fold to four-fold. As an exception, the n = 0 LLs are doubly degenerate due to the non-equivalence of the K and K$^{\prime}$ valleys. Particularly, the LLs at higher and lower energies remain split slightly. Such a LL splitting is stronger for higher $\Phi$. This points out that the TBLG with a small $\theta$ becomes more sensitive to magnetic field compared with other systems with a large $\theta$. Remarkably, the combination of Peierls substitution and tight-binding model employed in this work reveals perspicuous $B$ dependence of LLs at weak fields, which is quite different from the Hofstadter's butterfly by using the effective-mass approximation\,\cite{PRB2012Moon}. That is to say, our theoretical method allows one to explicitly analyze the vertical transition channels between occupied and unoccupied LLs. This is crucial for exploring the optical and transport properties of graphene-based materials.
\medskip

The probability distributions for four sublattices present well-defined oscillation modes, as observed from Fig.\,\ref{Fig12}(c). They are equivalent on $A$ and $B$ atoms. The LLs with the same zero modes, which are dominated by atoms on the first ($A^1$ and $B^1$) and second ($A^2$ and $B^2$) graphene layers, are separated by a narrow energy spacing. Generally, LLs having $n \geq 1$ zero modes are found possessing $n-1$ zero modes at the same time, which allows for the vertical transitions between the $n^{c,v}$ = $n,\,n \pm 1$ LLs. Consequently, the QHC exhibits the step structure of $4\,m$$e^2/h$, as presented in Fig.\,\ref{Fig12}(d). Especially, the $\pm 2 e^2/h$ steps near the zero energy, referring to the zoom-in of Fig. 12(d), correspond to the doubly degenerate n = 0 LLs. This feature is qualitatively similar to that of AB BLG, except that the plateaus here are three orders of magnitude smaller. We emphasize that the calculated magnetic-field-dependent LL spectrum and filling factors are in good agreement with the experimental results at the magic angle\,\cite{magic1, filling1, filling2}.
Note that, the filling factors of $\pm 1$ and $\pm 3$ reported in refs. \cite{filling1, filling2} might be induced by the LL splitting due to the effect of substrates. 
This implies the reliability of our proposed theoretical method.

\section{Concluding Remarks}
\label{sec4}

In a nutshell, we have proposed an approach by combining the generalized Peierls substitution with TBM to reliably calculate the magnetic quantization and QHE in TMLG systems. We have demonstrated that this method can efficiently solve the huge Hamiltonian matrix of TBLG and TDBLG in the presence of an applied magnetic field. For hetero-structures with large twist angles, the Bloch-basis TBM has been employed in order to reduce significantly the size of Hamiltonian matrix. We have further developed a simplified four-band TBM for the twisted system at the magic angle. By properly substituting the generalized Peierls phases into the TBM, we have successfully calculated the field-dependent LLs, the LL distribution probability, and the QHC of TMLG for both large and small twist angles. We emphasize that our theoretical prediction of LL spectra and filling factors are in good agreement with the experimental results.
\medskip

A notable accomplishment of this work is that we are able to provide incontrovertible explanations for the occurrence of QHC step structures by means of the unique selection rule for the inter-LL transitions based on analyzing the node structure of LL wave functions. This work opens up an opportunity for deciphering the interplay between an external magnetic field and the twisting effect on multilayer graphene. Remarkably, our proposed theoretical method can be used to understand the magnetic quantization of other complex systems, such as twisted graphene and substrates, other twisted hetero-structures, topological materials with both bulk and surface states. A deep understanding of QHC of various condensed matter systems is expected to be established.

\section*{Acknowledgement(s)}
P.-H. Shih would like to thank the Ministry of Science and Technology of Taiwan for the support through Grant No. MOST 110-2636-M-006-002. T.-N. Do would like to thank the NCKU 90 and Beyond project for the support. G.G. would like to acknowledge the support from the Air Force Research Laboratory (AFRL) through Grant No. FA9453-21-1-0046. D.H. would like to acknowledge the financial support from the Air Force Office of Scientific Research (AFOSR).

\end{document}